# Systems Theoretic Techniques for Modeling, Control and Decision Support in Complex Dynamic Systems

## Armen Bagdasaryan[*]

*Russian Academy of Sciences, Trapeznikov Institute for Control Sciences, 65 Profsoyuznaya, 117997, Moscow, Russia*

**Abstract:** Nowadays, modern complex systems of any interdisciplinary nature can hardly be analyzed and/or modeled without comprehensive usage of system theoretic approach. The complexity and uncertainty of the nature of modern systems, and the heterogeneity of related information, require a complex approach for their study, based on systems theory and systems analysis and consisting of information and expert knowledge management, initial pre-processing, modeling, simulation, and decision making support. As the complexity of systems increases, system theoretic methods become more crucial. Often they provide the only effective tools of obtaining the information about the elements in a system, connections between those elements, and the means for getting the adequate representation of system in a whole. The variety of complex systems can be described by deterministic or stochastic differential equations, statistical mechanics equations, neural network models, cellular automata, finite state machines, multi-agent systems, *etc*. Most of the complex real world objects are modeled as dynamic systems enriched by artificial intelligence resources. Equipped with artificial intelligence techniques, these models offer a wide variety of advantages such as coping with incomplete information and uncertainty, predicting system's behavior, reasoning on qualitative level, knowledge representation and modeling, where computer simulations and information systems play an important and active role, and facilitate the process of decision making.

This chapter aims to discuss the problems of modeling, control, and decision support in complex dynamic systems from a general system theoretic point of view, with special emphasis on methodological aspects. We consider the main characteristics of complex systems and of system approach to complex system study. Then the chapter continues with the general dynamic modeling and simulation technique for complex hierarchical systems functioning in control loop. The proposed technique is based on the information-mathematical models and described in terms of the hierarchical state transition diagrams. The methodology is sufficiently abstract to allow both qualitative and quantitative analysis of system state dynamics and control through hierarchical scenario calculus. The evaluation of different scenarios is defined by the multiple criteria vector-functions related to the efficiency of control strategies and time required for system goals achievement. We also offer general architectural and structural models of computer information system intended for simulation and decision support in complex systems.

## 1. INTRODUCTION

Complex System Science, as a field of research, has emerged in the past two decades. It is a multidisciplinary field aiming at understanding the complex phenomena of the real world that surrounds us. It studies how parts of a system give rise to the collective behaviors of the system and how the system interacts with its environment [1, 2].

The field of complex systems cuts across all traditional disciplines of science as well as physics, mathematics, biology, engineering, management, and medicine. It focuses on certain questions about parts, wholes and relationships. These questions are relevant to all traditional fields. Examples of complex systems are neural networks in the brain that produce intelligence and consciousness, artificial intelligence systems, swarm of software agents, social insect (animal) colonies, ecological and biological systems, traffic patterns, robotic systems, social and economic systems and many other scientific areas can be considered to fall into the realm of complex systems.

**[*]Address correspondence to Armen Bagdasaryan:** Russian Academy of Sciences, Trapeznikov Institute for Control Sciences, 65 Profsoyuznaya, 117997, Moscow, Russia; E-mail: abagdasari@hotmail.com

Complex systems are usually understood intuitively as a phenomenon consisting of a large number of elements organized in a multilevel hierarchical structure [3, 4], where elements themselves could represent systems (the concept "system of systems") [1]. The term complex is used to point out the fact that the problem treated here cannot be expressed only in quantitative relations but instead the most relevant values are qualitative.

So, the first and main characteristic of complex systems is that they contain a large number of mutually interacting entities (components, agents, processes, *etc.*) whose aggregate activity is nonlinear, cannot be derived from the direct summations of the activity of individual entities, and typically exhibit a some sort of self-organization (for example, hierarchical) [5-9]. Another important characteristic of complex systems is that the description of complex systems requires the notion of purpose, since the systems are generally purposive [2]. This means that the dynamics of the system has a definable objective or function. Each element of a complex system interacts with other elements, directly or indirectly. The actions of or changes in one element affect other elements. This makes the overall behavior of the system very hard to deduce from and/or to track in terms of the behavior of its parts. This occurs when there are many parts, and/or when there are many interactions between the parts. Since the behavior of the system depends on the elements interactions, an integrative system theoretic (top-down) approach seems more promising, as opposed to a reductionist (bottom-up) one.

Any scientific method (approach, technique) of studying complex real world systems relies on modeling (analytical, numerical) and computer simulation [10]. The study of complex systems begins from a set of models that capture aspects of the dynamics of simple or complex system. Most of the complex real world objects are modeled as dynamic systems [11]. These systems can be described by deterministic or stochastic differential equations, neural network models, cellular automata, finite state machines, multi-agent systems, *etc.* Most of the complex systems can be studied by using nonlinear mathematical models, statistical methods and computer modeling approaches. These models should be sufficiently general to encompass a wide range of possibilities but have sufficient structure to capture interesting features [12]. There are three interrelated approaches to the modern study of complex systems: (1) how interactions give rise to patterns of behavior, (2) understanding the ways of describing complex systems, and (3) the process of formation of complex systems through pattern formation and evolution.

But the final intention in the study of complex systems is to understand the real nature of the processes, their dynamics, their influence and interconnections, and the possible outcomes in order to make preventive actions and to make correct decisions. It also facilitates taking composite decisions, which are often the only possible ones in case of complex systems.

Moreover, as a rule, modern complex systems are large-scale. Large-scale systems are typically imposed a hierarchical structure in order to manage complexity. In hierarchical models, the notion of consistency is much important, as it ensures the implementation of high-level objectives by the lower level systems [3, 4, 8, 9, 12]. Such a description of system depends largely on the problem domain, specific goals and the point of view of the researcher. Although the problem of a researcher as an active element of the system has been considered in the literature to date, there is no unique opinion on the influence of the observer in the process of modeling.

Although many efforts have been made, there is no commonly accepted definition of a complex system. Heuristic approaches basically focus on the interaction between (microscopic) subsystems and the emergence of new qualities at the (macroscopic) system level, *e.g.*

- Complex systems are systems with multiple interacting components whose behavior cannot be simply inferred from the behavior of the components. - *New England Complex Systems Institute.*

- By complex system, it is meant a system comprised of a (usually large) number of (usually strongly) interacting entities, processes, or agents, the understanding of which requires the development, or the use of, new scientific tools, nonlinear models, out of equilibrium descriptions and computer simulations. - Journal *Advances in Complex Systems.*

Nevertheless, whatever definition one relies on, any complex system is a system with numerous components and interconnections, interactions or interdependencies which are difficult to describe, understand, predict, manage, design, and/or change [2, 12].

For this reason, computer simulations play a crucial role in studying complex systems and in understanding of how these systems function and work, and how they could be efficiently controlled.

Nowadays, information technologies and computer simulations have evolved into an essential tool for modeling, assessment and support in any domain requiring decision making. The complexity and uncertainty of the nature of complex systems, and the heterogeneity of related information, require a complex approach for their study, based on and consisting of data and knowledge management, modeling, simulation and, lastly, decision making support [13-15]. So, the search for the ways of formalization and automation of processes of modeling, control, and decision support in complex systems continues to attract much attention.

This chapter aims to discuss the problems of modeling, control, and decision support in complex dynamic systems from a general system theoretic point of view, with special emphasis on methodological aspects. We consider the main characteristics of complex systems and of system approach to complex systems study. The chapter continues with the general dynamic modeling and simulation technique for complex hierarchical systems consisting of many objects and functioning in control loop. The proposed technique is based on the information-mathematical models and is described in terms of the hierarchical state transition diagrams. The methodology is sufficiently abstract to allow both qualitative and quantitative analysis of system functioning and state dynamics through hierarchical scenario calculus. The evaluation of different scenarios is defined by the multiple criteria vector-functions related to the efficiency of control strategies and time required for system goals achievement. We also offer a general structure of computer information system intended for simulation and analysis of dynamic processes, control strategies and development scenarios in complex systems, and a structural scheme of decision support process.

The chapter is organized as follows. The next section presents the principles of system approach to complex systems study and the basic features of complex systems. The problem of modeling and control within the context of complexity of modern systems is addressed in section 3. The section 4 is devoted to the analysis of known existing paradigms and methods of mathematical modeling and simulation of complex systems, which support the processes of control and decision making. In section 5 we present the method of hierarchical state diagrams as a tool of dynamic modeling and simulation of complex hierarchical systems; we give conceptual principles of agent-based parametric modeling and then we describe the method in much detail. Then we outline some possible directions of further development of the proposed technique. The architecture of information system that supports simulation and analysis of dynamic processes and control scenarios, and decision making in complex hierarchical systems is proposed in section 6. In conclusion we make some final remarks on the topic of this chapter, and outline the areas of application of the presented technique.

## 2. SYSTEM THEORETIC APPROACH FOR COMPLEX SYSTEMS STUDY

The majority of real-life problems can be classified as complex ones, and, as a result, they inhabit some particular characteristics, which require interdisciplinary approaches for their study. Every complex system is an integration of interconnected parts and components (through informational, physical, mechanical, energetic exchange, *etc.*), which result in emerging of new properties and interaction with the environment as a whole entity. If some part is extracted from the system, it loses its particular characteristics and converts into an array of components or assemblies. An effective approach to complex system study has to follow the principles of system analysis [16-20], which are:

1. Description of the system. Identification of its main properties and parameters;

2. Study of interconnections amongst parts of the system, which include informational, physical, dynamical, temporal interactions, as well as the functionality of the parts within the system;

3. Study of the system interactions with the environment, in other words, with other systems, nature, *etc.*;

4. System decomposition and partitioning. Decomposition supposes the extraction of series of system parts, and partitioning suggests the extraction of parallel system parts. These methods can be based on cluster analysis (iterative process of integration of system elements into groups) or content analysis (system division into parts, based on physical partitioning or function analysis);

5. Study of each subsystem or system part, utilizing optimal corresponding tools (multidisciplinary approaches, problem-solving methods, expert advice, knowledge discovery tools, *etc.*);

6. Integration of the results received from the previous stage, and obtaining a pooled fused knowledge about the system. The synthesis of knowledge and composition of a whole model of the system can include formal methods for design, multi-criteria methods of optimization, decision-based and hierarchical design, artificial intelligence approaches, case-based reasoning, and others such as hybrid methods.

Basic reasons that make it difficult for complex systems to be described by formalized methods are the following ones:

- Information *incompleteness* on the state and the behavior of a complex system;

- Presence of a *human* (observer, researcher) as an intelligent subsystem that forms requirements and makes decisions in complex systems;

- *Uncertainty* (inconsistency, antagonism) and multiplicity of the purposes of a complex system, which are not given in a precise formulation;

- *Restrictions* imposed on the purposes (controls, behavior, final results) externally and/or internally in relation to a system are often unknown;

- *Weak structuredness*, uniqueness, combination of individual behaviors with collective ones are the intrinsic features of complex systems.

Complex systems are different from simple systems by their capabilities of:

- *Self-organization* - the ability of a complex system to autonomously change own behavior and structure in response to events and to environmental changes that affect the behavior.

For systems with a network structure, including hierarchical one, self-organization can amount to: (1) disconnecting certain constituent nodes from the system, (2) connecting previously disconnected nodes to the same or to other nodes, (3) acquiring new nodes, (4) discarding existing nodes, (5) acquiring new links, (6) discarding existing links, (7) removing or modifying existing links.

- *Co-evolution* - the ability of a complex system to autonomously change its behavior and structure in response to changes in the system environment and in turn to cause changes in the environment by its new (corrected) behavior.

Complex systems co-evolve with their environments: they are affected by the environment and they affect their environment.

- *Emergence* - the property that emerge from the interaction of constituent components of a complex system.

The emergent properties do not exist in the components and because they emerge from the unpredictable interaction of components they cannot be planned or designed.

- *Adaptation* - the ability of a complex system to autonomously adjust its behavior in response to the occurrence of events that affect its operation.

Complex systems should adapt quickly to unforeseen changes and/or unexpected events in the environment. Adaptation enables the system to modify itself and to revive in changing environment.

- *Anticipation* - the ability of a system to predict changes in the environment to cope with them, and adjust accordingly.

Anticipation prepares the system for changes before these occur and helps the system to adapt without it being perturbed.

- *Robustness* - the ability of a system to continue its functions in the face of perturbations.

Robustness allows the system to withstand perturbations and to keep its function and/or follow purposes, giving the system the possibility to adapt.

Being oriented on the analysis of complex object as a whole, the system approach does include the methods of decomposition of complex system on separate subsystems. But the main purpose is the subsequent synthesis of subsystems, which provides the priority of a whole. However, reaching this priority is not simple. For a number of complex systems, optimum of the whole system cannot be obtained from optimums of its subsystems. It should be noted, that complex systems that possess the property of integrity do not have constituent elements and act as one whole object. In this kind of systems, the connections and relations are so complicated and strong (all-to-all) that they cannot be considered as an interaction between the localized parts of system. In physical systems the integrity corresponds to locality that is to such an influence of one part of system to another which cannot be explained by interaction between them. As a rule, connections in integrative systems are often based on structural principles, but not on the cause/effect principle.

## 3. COMPLEXITY, MODELING, AND CONTROL IN COMPLEX SYSTEMS

Complex systems are more often understood as dynamical systems with complex, unpredictable behavior. Multidimensional systems, nonlinear systems or systems with chaotic behavior, adaptive systems, modern control systems, and also the systems, which dynamics depends on or is determined by human being(s), are the formal examples of complex systems [21-27]. In connection with modeling and control complexity, complex systems have specific characteristics, among which are:

- Uniqueness;

- Weak structuredness of knowledge about the system;

- The composite nature of system;

- Heterogeneity of elements composing the system;

- The ambiguity of factors affecting the system;

- Multivariation of system behavior;

- Multicriteria nature of estimations of system's properties; and, as a rule;

- High dimensionality of the system.

Under such conditions, the key problem of complex systems theory and control theory consists in the development of methods of qualitative analysis of the dynamics of such systems and in the construction of efficient control techniques. In a general case, the purpose of control is to bring the system to a point of its phase space which corresponds to maximal or minimal value of the chosen efficiency criterion. Another main and actual problems in the theory of complex systems and control sciences is a solution of "ill-posed, weakly- and poorly-structured and weakly-formalized complex problems" associated with complex technical, organizational, social, economic, cognitive and many other objects, and with the perspectives of their evolution. Since the analysis and efficient control are impossible without a formal model of the system, the technologies for building the models of complex systems have to be used.

Complexity of a system is a property stipulated by an internal law of the system that defines some important parameters, including spatial structure and properties of the processes in this structure. This definition of complexity is understood as certain physical characteristic of nature. Since it is a nonlinearity of internal regularities (laws) that underlies the complexity of real world systems, complexity and nonlinearity are sometimes considered as synonyms. And the more complex a process or geometrical form of a system (or object) is, the more it is nonlinear.

Complexity is a many-faceted concept. Today we can distinguish several basic forms of complexity: structural, geometrical, topological, dynamical, hierarchical, and algorithmic. However, other possible forms of complexity can be found as well. For example, one that comes from large scales.

Large-scale control systems typically possess a hierarchical architecture in order to manage complexity. Higher levels of the hierarchy utilize coarser model of the system, resulting from aggregating the detailed lower level models. In this layered control paradigm, the notion of hierarchical consistency is important, as it ensures the implementation of high-level objectives by the lower level systems [28-30]. Large-scale systems are systems of very high complexity. Complexity is typically reduced by imposing a hierarchical structure on the system architecture [3, 4, 6-9, 12]. Hierarchical structures for discrete event systems have been considered in multiple works [30-33]. In such a structure, systems of higher functionality reside at higher levels of the hierarchy and are therefore unaware of unnecessary lower-level details. One of the main challenges in hierarchical systems is the extraction of a hierarchy of models at various levels of abstraction which are compatible with the functionality and objectives of each layer. The notions of abstraction or aggregation refer to grouping the system states or control objects into equivalence classes [30].

Algorithmic complexity finds itself in many software systems. These are the most complex systems developed by human being, although their structure and dynamics are comparatively simple. Structural, dynamical, algorithmic, hierarchical and large-scale complexities of systems attract much of attention because we face them, manifestations of nonlinearity of nature, in our everyday life.

Interplay between intellectualized mathematical and information technologies of control and decision support plays an important role in modeling of processes of evolution and functioning of complex systems. Intellectualization of complex control systems has actively been developing in recent decade. In order to intellectualize modern control systems, the artificial intelligence methods or intelligent subsystems embedded in control system are more often applied [34, 35, 46, 104]. The intellectualization of complex systems seems to be a positive and very perspective direction of control systems development. It significantly eases control decision making, as the underlying mechanisms are similar to those used by human intellect. Control processes in intellectualized systems are based on the experience, skills and knowledge, that is, they are mostly based on the "understanding" of complex situations of purposeful behavior. An elementary intellect in control systems is constructed by using feedback loops and information flows, which give a system the capability of "understanding" of current situations. To "understand" the more complex situations, an adaptive subsystem should be added to the main feedback loop. However, if the object/system performs multiple functions (multipurpose object) then each function makes the system more complex and, as a consequence, the more sophisticated intelligent subsystems have to be used. But many questions remain: how to differ control systems by the level of intelligence; at what level of evolution of control systems they can be considered as artificial intelligence systems; what is the relation between adaptive properties and intelligent systems.

For modeling and analysis of complex control systems in the presence of principally non-formalizable problems and impossibility of strict mathematical formulation of problems, expert knowledge and information databases are used. Construction of models of complex systems is accompanied by extensive use of expert knowledge and information about the system stored in data- and information systems. This knowledge should be integrated in a unified way. Qualitative character of most of parameters of complex systems results in knowledge fuzziness and uncertainty and, as a consequence, in problem of its formalization.

The analysis above supports the fact that complex systems are usually difficult to model, design, and control. There are several particular methods for coping with complexity and building complex systems. At the beginning, a conceptual model of system is developed, which reflects the most important, in the context of the problem under study, material and energy and information processes taking place between different elements of system (or, subsystems), internal states of which can be considered as independent. This kind of model determines the general structure of system and it should be complemented by algorithmic and, more often, by mathematical models of each of the subsystems. These models can be represented by graph models, Petri nets models, system dynamics models or by their combination. The obtained models are aggregative that reflect the dynamics of the most important, for the current investigation, variables. Then, the next step consists in checking the mathematical models for their behavioral adequacy to real system, and in identification of parameters of the models over the sets of admissible external actions and initial/boundary conditions. The difficulties of solution of these problems increase as the system becomes more complex. For this reason, another important step is the structuring of problem domain (or situations), control domain, and simulation scenarios. For these purposes, stratified models, state (or flow) diagram models, system dynamics models, aggregative models and robust identification can be used. However, the developed model should be subjected to intense analysis and possible changes after its testing for structural controllability, observability, identifiability, and sensitivity. These properties guarantee the model rigidity in a given class of variations of the problem conditions and, as a consequence, the reliability and accuracy of system simulation. Besides that, the rigidity enables one to reduce the model to canonical (more simple) forms, which leads to significant simplification of modeling, control synthesis, and analysis of the system. Thus, when constructing a model of complex dynamical system, three forms of its description arise: (1) conceptual model, (2) formalized model, (3) mathematical model, and (4) computer model.

## 4. MATHEMATICAL AND SIMULATION MODELING OF COMPLEX DYNAMIC SYSTEMS: EXISTING FRAMEWORKS AND PARADIGMS

The system approach assumes that any object under study is considered as a single complex system together with the control subsystems. To ensure high quality management one should be familiar with the properties of controlled subsystems. In order to identify the properties of controlled subsystems, as well as to evaluate the quality of decisions, their responses to applied decisions and activities, and the results of automated modeling of the subsystems in different predictable conditions should be used in the process of functioning of the object.

In systems simulation there are several paradigms, formulations of problems, and approaches to their solution, which are used as a "framework" for building and analyzing models [36, 37]. There are several quite different viewpoints on system modeling: dynamical systems, system dynamics, discrete event systems, cellular automata models, neural network models, finite state machines, cognitive modeling (cognitive maps), and multi-agent models.These paradigms differ, rather, by concepts and views on the problems and approaches to solving them, than the applications areas. Often, the adherents of a paradigm believe that the right formulation and solution of problems of modeling and simulation can only be possible within the framework of concepts and techniques of this particular paradigm. For example, the advocates of modeling and analysis of dynamical systems believe that other approaches are "not entirely" scientific, or they are just special cases of system representation and analysis as systems of algebraic differential and/or integral equations. In fact, each of the paradigms has their own right for a life, and use of one or another paradigm is determined only by the aims of modeling and is associated with those aims by means of the chosen level of abstraction for solving (control) problems.

### 4.1. Aggregative Models of Systems

Aggregative model describes the control object as a multilevel structure of dynamical systems of given types, or as aggregates [38]. In this description, the system is viewed as a synthesis (the most general and most complex) of a generalizing class of complex systems, and is called aggregative system [38-40]. The aggregate is used to model the elementary blocks of complex system. So, aggregative system is a system composed of any set of aggregates, if the transmission of information between them is assumed to be instantaneous and without distortion.

Let $T \subset R$ be a subset of real numbers, $X, U, Y, Z$ be sets of any nature. The elements of these sets are interpreted as follows: $t \in T$ is a time instant, $x \in X$ is an input signal, $u \in U$ is a control symbol (signal), $y \in Y$ is an output signal, $z \in Z$ is a state. States, input, control, and output signals are considered as functions of time, $z(t), x(t), u(t)$, and $y(t)$. Formally, aggregate is a mathematical model represented by the tuple

$$A = \{T, X, U, Y, Z, H, Q\}$$

where $H$ and $Q$ are operators (in general case, random ones); $H$ is a transition operator which defines the current state on the basis of previous states (history); $Q$ is an output operator. These operators realize the functions $z(t)$ and $y(t)$. In general, all sequences of events in the aggregate are realizations of random sequences with the given distribution laws. The structure of these operators distinguishes the aggregates amongst any other systems.

As a rule, it is always possible to distinguish two types of states: (1) ordinary states in which the system remains almost all the time, (2) special states, specific to the system in some isolated time instants, coinciding with the receipt of input and control signals or with the issuance of the output signal; at these moments the aggregate may change its state discontinuously, but between the special states the change of the coordinates occurs smoothly and continuously.

The aggregate is a mathematical scheme of quite a general type, special cases of which are Boolean algebras, contact relay networks, finite automata, dynamical systems, described by ordinary differential equations, and some other mathematical objects.

The model of aggregate can be used as a model for a discrete-continuous system (hybrid system) in a whole or as a basic model for its components. In the latter case, the system is represented as a network of aggregates with fixed communication channels. As an example of system of this kind, a complex system of aircraft flight control of a large airport can be considered; it is as the automated control system with extensive and large distributed information system and with very complex information processing algorithms.

**4.2. Discrete Event Models and Hybrid Systems**

Discrete event approach [41] to modeling of complex systems can be used for formal description of both discrete and discrete-continuous (hybrid) systems. The description formalism includes mathematical model of system, specification language, as well as a set of procedures and functions that implement the simulation algorithm [42]. In contrast to the aggregative approach, the simulation algorithm is based on the discrete event dynamic modeling of complex systems [43-46].

The discrete-continuous system is a mathematical model of the form

$$S = \{T, P, e, E, K, F\}$$

where $T = \{t_i\}$, $t_i \in R \geq 0$ is a discrete model of time;

$P$ is a set of process classes;

$e$ is a set of event classes (the causes of instantaneous change of behavior and structure of the system);

$E$ is a set of algorithms for the event classes (preparatory discrete actions on the transition to new behavior of the system);

$K$ is a calendar of event planning;

$F$ is a set of equations that describe the local behavior of system processes at time intervals between the events.

The calendar of event planning contains the information about the events in different objects; the dynamics of system is described with the help of the calendar.

The model of calendar of event planning can be given as the tuple

$$\boldsymbol{K = \{\langle t_i, e_i \rangle, L\}}$$

where $L$ is a condition for occurrence of the event (planning of event by condition, it can be given in the form of logical formula, predicate).

The structure and behavior of a process is described by the mathematical model

$$P = \{X, Y, V_s, V_d, B\}$$

where $X, Y$ are input and output channels;

$V_s$ is a set of static process variables that are defined by algebraic expressions and can be changed only upon the implementation of event algorithms;

$V_d$ is a set of functions, dynamic variables, that are defined by differential equations from $F$;

$B$ is a main body of the process that consists of the descriptions of its various behaviors.

Modeling the behavior of discrete-continuous system means building of a set of event sequences leading to the change of system's behavior and structure, assuming the initial state to be an event. The global behavior is modeled using a special monitor process which shifts the system time in accordance with the calendar of event planning, or in accordance with the analysis of the occurrence time of the event, which is scheduled by the condition. The simulation process ends when the event calendar is empty.

As we can see, the aggregative models and discrete event models use different approaches to the description of the same class of complex systems. One can note that almost all the approaches to modeling can be considered as an extension of the basic (discrete or continuous) model, so the behavior of discrete-continuous system is represented in them from different viewpoints and from different, sometimes contradictory, emphases.

The "hybrid" line of study of discrete-continuous systems emerged in the early 90's. A. Pnueli and D. Harel, the founders of hybrid approach, introduced a new class of complex systems, hybrid reactive systems. Investigation of the behavior of the hybrid system reduces to static qualitative analysis of behavioral properties, without the use of point-wise numerical simulation of the global system behavior. Several software tools have been developed for modeling hybrid systems, for example, HyTech is an automatic tool for the analysis of embedded systems.

**4.3. System Dynamics Model**

For the analysis of complex systems with nonlinear feedbacks, the Forrester's system dynamics is used [47]. The method of system dynamics is employed for modeling systems from various backgrounds [48-50]. System Dynamics as a method of simulation includes the following stages:

- Structuring of an object;

- Building a system diagram of the object, where the links between the elements are specified;

- Determining system variables for each element and the rate of their growth;

- Adopting the hypotheses about the dependence of each rate of growth on the variables, and formal description of these hypotheses;

- Evaluating the process for the parameters introduced with the use of available statistics.

To construct and study models with system dynamics method, a specialized programming language Dynamo that combines simulation tools and graphical notations was developed.

The Forrester's model includes the following components:

- Levels (resources);

- Flows that move the content of one level to another level;

- Decision Functions that govern the rate of flows between the levels;

- Information channels that connect the above functions with the levels.

Levels characterize the resulting accumulations within the system. Each level is described by its variable that depends on the difference between the incoming and outgoing flows. The rates determine the instantaneous current flows between the levels in the system and model the jobs, while the levels measure the state which the system reaches as a result of some job.

Decision functions and equations of rates formulate the rules of behavior that determine how the available information about the levels leads to the choice of decisions related to the values of the current rates. The basic structure of the model, presented in Fig. (**1**), shows only one network with the elementary scheme of information links between levels and rates. However, to model the activities, for example, of a whole industrial enterprise, one needs to construct multiple interconnected networks.

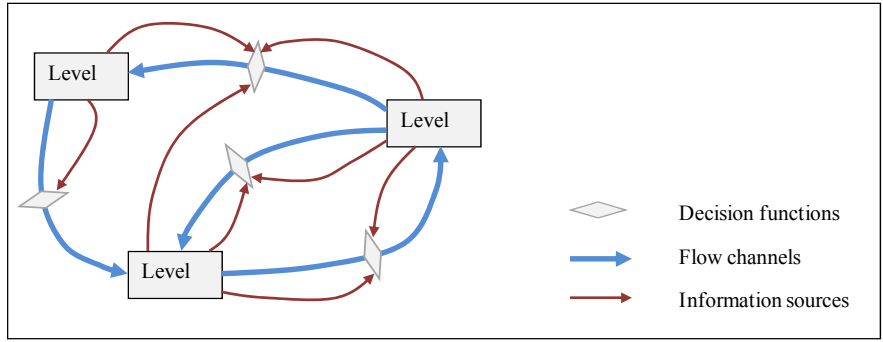

**Figure 1:** Basic Structure of Forrester's Model.

There are six types of networks which represent substantially different types of variables: orders, materials, finances, manpower and equipment, connected together through the information network. Each of these networks can further be broken down into several separate parts.

Information network serves as a connective tissue for other types of networks. It sends the information from the levels to the decision points, as well as the information about the rates in two other networks to the levels of information network. There exist levels and rates in the information network as well. For instance, information on the actual current rate in the flow of materials is averaged to determine the level of the average rate of flow of materials. This level refers to the information network.

The basic structure of the model is supplemented by a system of equations, which connect the characteristics of the levels of this structure. Basically this system consists of the equations of two types: the equations of levels and the rate equations.

When constructing the equations, the time axis is divided into intervals $\Delta \tau_{ij}$ between $i$-th and $j$-th time instants. The new values of levels are calculated at the end of the interval, and they are used to determine the new rates (decisions) for the next interval $\Delta \tau_{ij+1}$.

The level equations have the form

$$Q_{ij} = Q_{ik} + \Delta \tau_{kj} \left( \sum_{x'} q_{x_1,i}^{k,j} - \sum_{x''} q_{x_2,i}^{k,j} \right)$$

where $x_1 \in x'$, $x_2 \in x''$, $x' \cup x'' = x$, $x$ is the set of levels related to the $i$-th level;

$Q_{ij}$ is the value of $i$-th level at $j$-th time instant;

$\Delta \tau_{kj}$ is the length of the interval from time instant $k$ to time instant $j$;

$q_{x_1,i}^{k,j}$ is the rate of input flows for $i$-th level at the interval between time instant $k$ and $j$;

$q_{x_2,i}^{k,j}$ is the rate of output flows for $i$-th level at the interval between time instant $k$ and $j$.

The rate equation has the form

$$q_j^{k,j} = S_k / \Delta t$$

where $S_k$ is the value of level that determines the delay at time instant $k$;

$\Delta t$ is a constant, mean time, required to overcome the delay.

The stages of building of the model are the following:

1. Construction of the basic structure of the model as a specialized graph;

2. Parameterization of the graph and the construction of the corresponding system of equations;

3. Description of the model, using a simulation language, and conducting experiments.

The advantages of the model include: (1) the ability to reflect almost any causal relationships, (2) simple mathematical formulation, and (3) the use of terminology, synonymous to the concepts of the languages of economy, automation, and engineering.

For the main phase variables, the so-called system levels, the following differential equations, the same for all variables, are used.

$$\frac{dy}{dt} = y^+ - y^-$$

where $y^+$ is the positive rate of variable $y$, which includes all the factors that cause its growth;

$y^-$ is the negative rate of variable $y$, which includes all the factors that cause the decrease of $y$.

These rates are supposed to be expressed in terms of the product functions that depend only on the so-called "factors", the auxiliary variables that are combinations of the main variables:

$$y^\pm = g(y_1, y_2, \ldots, y_n) = f(F_1, F_2, \ldots, F_k) = f_1(F_1) f_2(F_2) \ldots f_k(F_k)$$

where $F_i = g_i(y_{i_1}, \ldots, y_{i_m})$ are factors; $m = m(j) < n$, $k < n$, $n$ is the number of levels, that is, the number of factors is less than that of variables which allows one to simplify the problem and consider only the functions of one variable.

The software tools that implement system dynamics models and, in particular, the model of Forrester are the well-known simulation environments, such as AnyLogic, VenSim, PowerSim, ModelMaker, *etc*. To construct the models they use a graphical representation of the dependent variables in the form of the so-called "stock and flow diagrams".

**4.4. Cellular Automata**

Cellular automata [51-53] have played a considerable role as simple models for the study of complex systems, and have recently been popularized by Steven Wolfram in his well-publicized book "A New Kind of Science" [54, 55] and by John Conway's model Game of Life. Cellular automata were originally invented and studied by the mathematicians Stanislaw Ulam and John von Neumann. Von Neumann introduced them as a biologically motivated computation models, and adopted cellular automata as a substrate for constructing a self-reproducing automaton [56].

A cellular automaton is a decentralized model providing a platform for performing complex computations with the help of only local information. The cellular automata paradigm of local information, decentralized control and universal model of computations is exploited by many researchers and practitioners from different fields for modeling various applied systems [57-69]. Cellular automata, both deterministic and stochastic, are used for modeling of various complex phenomena in different branches, from physical to social to engineering ones [58, 59, 68, 69].

The popularity of cellular automata can be explained by their simplicity, and by the enormous potential they hold in modeling complex systems, in spite of their simplicity. Cellular automata can be viewed as a simple model of a spatially extended decentralized system made up of a number of individual components (cells). The communication between constituent cells is limited to local interaction. Each individual cell is in a specific state which changes over time depending on the states of its local neighbors. The overall behavior can be viewed as a simultaneous change of states of each individual cell.

A cellular automaton consists of a number of cells organized in the form of a lattice. It evolves in discrete space and time, and can be viewed as an autonomous finite state machine. A cellular automaton is typically defined over a $d$-dimensional integer regular lattice $\mathbf{Z}^d$. Because of its inherent simplicity, the one-dimensional cellular automaton, $d = 1$, with two states per cell became the most studied variant of cellular automaton.

More often, a cellular automaton is defined over a two-dimensional lattice such as $\mathbf{Z}^2$. Each lattice point $(x_i, x_j)$ is referred to as a cell, site, or node, and is denoted by $x = (x_i, x_j)$. Each cell has a state $s_x(t)$ which often takes on its values from $F_2 = \{0, 1\}$, that is, each cell stores a discrete variable at time $t$ that refers to the present state of the cell. For each cell, a notion of a neighborhood is defined. The neighborhood $N_x$ of a cell $x$ is the collection of cells that can influence the future state of the given cell (Fig. (**2**)).

A common choice is the Moore neighborhood, which is the $(2\lambda + 1) \times (2\lambda + 1)$ square block centered at the cell $x = (x_i, x_j)$, where $\lambda$ is a positive integer parameter called the range. More precisely, it is the set

$$\boldsymbol{N_x = \{y = (y', y''): |x_i - y'| \leq \lambda \text{ and } |x_j - y''| \leq \lambda\}}$$

Another choice in the two-dimensional case is the von Neumann neighborhood defined as

$$\boldsymbol{N_x = \{y = (y', y''): |x_i - y'| + |x_j - y''| \leq \lambda\}}$$

The neighborhood generally varies from three to five or seven cells.

The next state of the cell at $(t + 1)$ is affected by its state and the states of its neighbors at time $t$. Based on its current state $s_x(t)$ and the current states of the sites in its neighborhood $N_x$, a function $f_x$ (called next state function, update rules, or local transition function) is used to compute the next state $s_x(t + 1)$ of the cell $x$.

That is, we have the equation

$$\boldsymbol{s_x(t + 1) = f_x(\tilde{s}_x(t))}$$

where $\tilde{s}_x(t)$ denotes the set consisting of all the states $s_{x'}(t)$ such that $x' \in N_x$.

The configuration $s(t)$ of a cellular automaton is the tuple consisting of the states of all the cells. The above equation is used to map the configuration $s(t)$ to $s(t + 1)$. The cellular automaton dynamics or the dynamical system defined by the cellular automaton is the map $\Phi$ that sends $s(t)$ to $s(t + 1)$. The study of cellular automaton consists mainly in understanding its evolution, how configurations evolve, under the repeated iteration of the map $\Phi$, so the dynamical behavior is generated and then can be studied.

The local rules applied to each cell can be either identical or different. These two different possibilities are termed as uniform and hybrid cellular automaton respectively.

While the next state function (local rule) in general is deterministic in nature, there are variations in which the rule sets are probabilistic [64, 66], or fuzzy [68, 69]. The nature of next state functions also varies significantly; researchers have defined the rule set according to the design requirements of the applications. In another type of cellular automata, the states are assumed to be a string of elements in a Galois field $GF(q)$, where $q$ is the number of states of a cellular automaton cell.

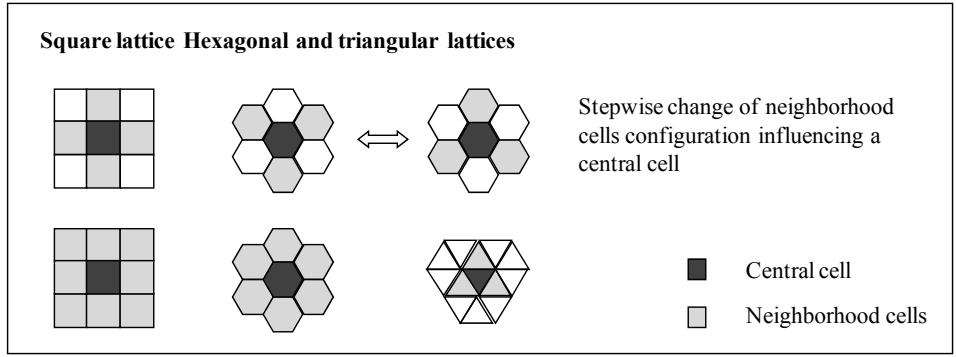

**Figure 2:** Types of neighborhoods in cellular automata.

### 4.5. Neural Networks

The process of complex systems modeling requires a large amount of knowledge about the objects, including experimental, monitoring, and expert information. The last decades are witnessed with the wide use of neural networks to cope with large amounts of data and for processing of information [70, 71]. Neural network based modeling has various application areas, such as pattern recognition, adaptive control, functional approximation, prediction, expert systems, the organization of associative memory, as well as study and analysis of technical and engineering systems and automation processes, and other applications. Each neural network consists of a collection of neurons (Fig. (**3**)), which are simple elements of information transformation.

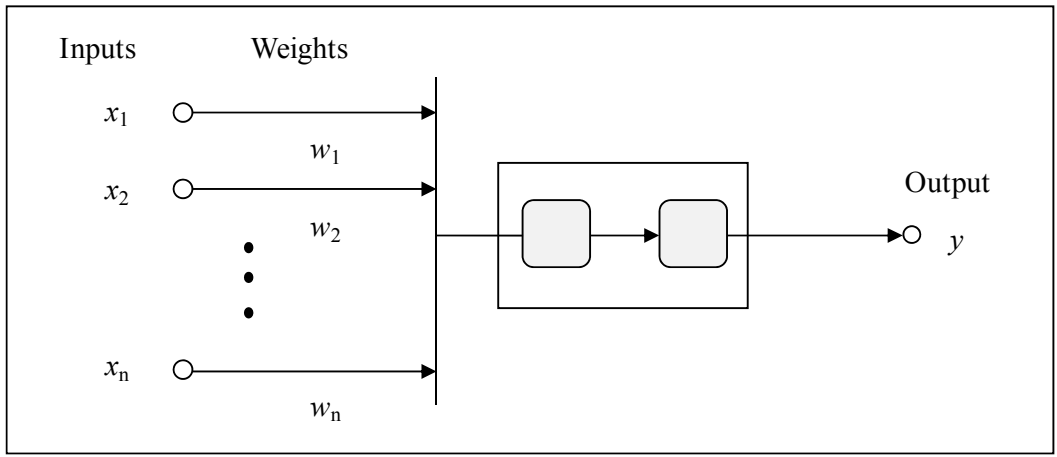

**Figure 3:** Model of artificial neuron.

The most well-known model of neural network is the Hopfield network [72]. Hopfield networks were proposed as a model of associative memories. Hopfield network is a recurrent artificial neural network constructed from artificial neurons; it is a network of some fixed numbers of such artificial neurons, which are in most cases fully connected.

A discrete Hopfield neural network consists of an undirected graph $G(V, E)$. At any time $t$, each node $v_i \in V$ has a state $x_i(t) \in \{+1, -1\}$. Each node $v_i \in V$ has an associated threshold $\tau_i \in \mathbf{R}$. Each edge $\{v_i, v_j\} \in E$ has an associated weight $w_{ij} \in \mathbf{R}$. For each node $v_i$, the neighborhood $N_i$ of $v_i$ is defined, which includes $v_i$ itself and the set of nodes that are adjacent to $v_i$ in $G$. More formally,

$$N_i = \{v_i\} \cup \{v_j \in V : \{v_i, v_j\} \in E\}$$

States of nodes are updated as follows. At time $t$, node $v_i$ computes the function $f_i$ defined by

$$f_i(t) = \text{sgn}\left(-\tau_i + \sum_{v_j \in N_i} w_{ij} x_j(t)\right)$$

where sgn is the map from **R** to $\{+1, -1\}$, sgn : **R** $\to \{+1, -1\}$, defined by

$$\text{sgn}(x) = \begin{cases} +1, \text{if } x \geq 0 \\ -1, \text{otherwise} \end{cases}$$

Now, the state of $v_i$ at time $t + 1$ is

$$x_i(t + 1) = f_i(t)$$

It is assumed in many works that the underlying undirected graph is complete, that is there is an edge between every pair of nodes. In the definition presented above, the graph need not be complete. However, this does not cause any difficulties since the missing edges can be assigned weight 0. As a consequence, such edges will not play any role in determining the dynamics of the system.

So, given the weights, thresholds, and the updating rule for nodes the dynamics of the network is defined if we determine in which order the nodes are updated. There are two ways of updating them: both synchronous and asynchronous update models of Hopfield networks have been considered in the literature.

- *Asynchronous*: one fix one node, calculates the weighted input sum and updates immediately. This can be done in a fixed order, or nodes can be picked at random, which is called asynchronous random updating.

- *Synchronous*: the weighted input sums of all nodes are calculated without updating the nodes. Then all nodes are set to their new value, according to the value of their weighted input sum.

Neural networks are related to the class of informational models. There are several types of informational models based on neural networks that are presented in the literature:

- Modeling a system in response to external action;

- Classification of internal states of a system;

- Prediction of dynamics of changes in a system;

- Assessment of completeness of description of a system and determination of the comparative importance of system parameters;

- Optimization of system parameters with respect to a given objective function;

- System control.

Typically, informational models are less expressive compared with formal mathematical models and expert systems by the criterion of "explicability" of granted results, but the lack of restrictions on the complexity of systems simulated determines their important practical significance. However, in some cases, neural networks and mathematical models can be combined in one model, for example, when external conditions are described by the equations of mathematical physics, but the response of a system is modeled by neural network. Sometimes, hybrid neural networks, which have fuzzy parameters [73-75], are used.

Synthetic models, which are based on a synthesis of system components, are practically the only alternative modeling tools in the area of complex systems and their control, examples are sociology, long-term weather forecasting, macro-economy, and medicine. Recently, synthetic information models, which include neural networks, have become widely used in the modeling and study of technical, engineering, and automation systems.

**4.6. Cognitive Analysis and Cognitive Maps**

Among the control problems the most difficult are those connected with the solution of complex problems. The aim of these kinds of problems is to shift the situation in a problem domain to a desired direction. In this case, the control object is the whole problem domain, which is regarded as a dynamic situation, consisting of a set of heterogeneous interacting factors. Some of these factors directly depend on the decisions of Decision-Maker (DM); other factors depend on DM only indirectly, while others do not depend at all and are considered as external disturbances. Dynamics of the situation is reflected in the fact that the situation changes with time under the influence of DM actions, external disturbances, and under the effects of some factors on others.

As a rule, when solving this kind of complex problems, one evidently faces the fact that, unlike most of the technical and engineering systems, the control objects (*i.e.*, situations) are both weakly-formalized and ill-structured:

- The system of concepts (factors) and connections between them is not defined with sufficient degree of completeness;

- The basic parameters of the situation (values of the factors, degree of influence of some factors on the others) are mostly not quantitative but have qualitative characteristics - numbers, intervals, fuzzy values, linguistic estimates that form a linearly ordered scale;

- Non-stationarity of the processes themselves, and the variation of certain characteristics of the processes is often unknown, making it difficult to build their quantitative models;

- The values of parameters of the situation, factors and connection strength are obtained mainly not on the basis of objective measurements but they are based on expert knowledge and estimates, which are subjective opinions;

- The alternatives of how the situation will evolve cannot be formulated beforehand; they arise in the process of the situation analysis.

The above features restrict the capabilities of simulation modeling, oriented on the use of quantitative characteristics, and of methods of traditional decision theory which relies on the methods of selecting the best possible alternative from a set of well-formulated alternatives. For this reason, these approaches became less efficient when applied to decision support and modeling in ill-structured problem domains.

The modern approach to the modeling, control, and analysis of ill-structured problem domains and complex control problems, is based on the notion of cognitive map [76, 77], and is called cognitive analysis of situations. Cognitive map is a model of representation of expert knowledge about the situation under study. Cognitive maps are used to describe causal relationships and influences between factors, as well as to model the dynamics of weakly-formalized systems [78-83].

Mainly, cognitive approach, based on cognitive aspects of model construction, includes the processes of perception, thinking, knowledge, explanation and understanding. To picture this, the schematic, simplified description of the world, related to the problematic situation, is portrayed as a cognitive map. From the standpoint of cognitive approach the process of modeling can be divided into several stages as represented in Fig. (**4**).

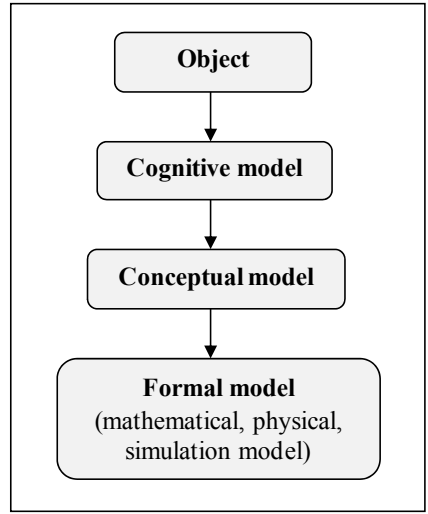

**Figure 4:** Generalized scheme of the process of modeling based on cognitive approach.

The main purpose of the use of cognitive maps is a qualitative analysis and modeling of the dynamics of situations (tendencies, directions of changes of the values of factors, the study of scenarios, *etc*.). For the purposes of quantitative analysis, the theory of differential or difference equations and optimal control theory are traditionally used, and in case of game-theoretic setting the theory of dynamic or differential games is applied (see Table **1**).

**Table 1:** Methods for qualitative and quantitative analysis of dynamics of situations

| Purpose | Modeling Method | |
| --- | --- | --- |
| | *Qualitative analysis* | *Quantitative analysis* |
| Description of situation | Cognitive maps | Differential or difference equations |
| Analysis and control of situation | Simulation modeling | Optimal control theory |
| Analysis of interrelationship of agents interested in the development of situation | Cognitive games | Dynamical games |

The mathematical model of directed graph (digraph), and its extensions, underlies the notion of cognitive map. The mathematical model of digraph is extended to get the mathematical models of signed, weighted signed and functional signed digraphs.

The model of digraph $G(V, E)$ is extended by the components:

- $A = \{A_i;\ i \leq n, n = |V|\}$; each vertex $v_i \in V$ is assigned a parameter $A_i \in A$;

- $F(A, E)$ - a functional that determines the arcs transformation; each arc is corresponded with either a sign or a weight or a function.

If the functional $F$ has the form

$$F(A_i, A_j;\ e_{ij}) = \begin{cases} +1, \text{if an increase (decrease)of } A_i \text{ causes} \\ \quad\quad \text{the increase (decrease)of } A_j \\ -1, \text{if an increase (decrease)of } A_i \text{ causes} \\ \quad\quad \text{the decrease (increase)of } A_j \end{cases}$$

then the model is called a signed digraph.

If the functional $F$ has the form

$$F(A_i, A_j; e_{ij}) = \begin{cases} +w_{ij}, \text{if an increase (decrease) of } A_i \text{ causes} \\ \quad\quad \text{the increase (decrease) of } A_j \\ -w_{ij}, \text{if an increase (decrease) of } A_i \text{ causes} \\ \quad\quad \text{the decrease (increase) of } A_j \end{cases}$$

then the model is called a weighted signed digraph; $w_{ij}$ is called a weight of the arc.

If the functional $F$ has the form

$$F(A_i, A_j; e_{ij}) = f_{ij}(A_i, A_j)$$

then the model is called a functional signed digraph.

The choice of concrete model depends largely on the problems of analysis to be addressed in the process of usage of the model. The problems of analysis of situations based on the cognitive maps can be divided into two types: static and dynamic. There are two kinds of models that correspond to these types: (1) static analysis, or analysis of influences, is the analysis of the current situation that involves the separation and comparison of causal chains, paths of influence of some factors on the others through the third ones, *i.e.*, indirect influence. The aim of dynamic analysis is to generate and analyze possible situation development scenarios at time scales. In both cases, the purpose of the analysis consists in the formation of possible alternative control decisions. These alternatives are the sets of control factors, *i.e.* factors, the change of which can be directly influenced by decision maker.

When setting control problems on cognitive maps, the following concepts are often used. Control factors are those factors which the decision maker can change. The purpose of control is to achieve certain values of some selected factors, which are called the target factors. External, or input, factors are the factors that are not influenced by the other factors of a cognitive map.

Analysis of influences identifies the factors with the strongest influence on target factors, that is, it determines the most effective points for applying control actions. Dynamic analysis considers the development of situation as a process of changes of its states in discrete time, and under the state of the situation $S(t)$ at time $t$ the set of values of all its factors $(A_1(t), ..., A_n(t))$ at this time moment is understood. Dynamic analysis solves two main problems. The direct problem is a prediction of the situation development for the given control or external actions (change in the values of some control or external factors), that is, it is the computation of the sequence $X(1), X(2), ..., X(n)$ for a given change of the state $X(0)$. The inverse problem is the calculation of control actions that result in the situation having a given state, for example, the state where the target factors have the desired value or some close to it.

We can now distinguish several main directions of control of ill-structured situations on the basis of cognitive models:

- Developing methods of analysis of cognitive maps and decision support techniques based on models of cognitive maps;

- Developing the techniques of problem domain structuring, in other words, the methods for construction of cognitive maps;

- Developing software tools and information technologies that implement the above methods.

By interpreting the vertices, arcs and weights, and different functions that define the influence of connections on factors, in various ways, one can arrive at different models of cognitive maps and at different means of their analysis. The most common classes of models are signed digraphs, weighted cognitive maps, linear models, and fuzzy models.

**Signed digraphs.** According to the above definitions, one considers a digraph $G(V, E)$ and obtains a signed digraph if there is a function of sign: $E(G) \to \{-1, +1\}$. The weight sign $+1$ means the positive influence, the sign $-1$ means negative influence. The weight of a path is the product of weights of its arcs; it is positive if the number of negative arcs is even, and is negative if this number is odd. If there are both positive and negative paths between vertices $v_i$ and $v_j$, then the character of the total influence of factor $v_i$ on factor $v_j$ remains uncertain. The computation of influences can be defined as follows.

*Indirect influence* $I_p$ of factor $v_i$ on factor $v_j$ *via* the path $P$ from $v_i$ to $v_j$ is determined as

$$I_p = \prod_{(k,l) \in E(P)} w_{kl}$$

where $E(P)$ is the set of arcs of the path $P$, and $w_{kl}$ is the weight (sign) of the arc $e_{kl} \in P$.

The total influence $T(i, j)$ equals $+1$ if all $I_p > 0$, and is equal to $-1$ if all $I_p < 0$.

The analysis of cycles in signed digraph is one of the problems that have to be solved in the process of its analysis. The positive cycle is a positive feedback loop; for example, if the factors are assigned some values, then an increase of the value of a factor in cycle leads to its further increase and then to its unbounded growth, which, in turn, leads to the loss of stability. The negative cycle counteracts the deviations from the initial state and contributes to stability; however, some instability in the form of significant fluctuations occurring during the passage of excitation along the cycle is still possible. Structural analysis and study of structural properties of signed digraphs, development of heuristic algorithms and estimations of significance of the elements of signed digraph, verification of weights of arcs, and estimates of preference degree of factors are the main analytical problems in the area.

The main disadvantage of signed models is the impossibility to cope with the strengths of influences over different arcs and paths, as well as the lack of a mechanism to resolve the uncertainties that arise in case of both positive and negative paths between two vertices. The basic approach to address them is to introduce weights that characterize the strength of influence, which leads to the weighted signed (cognitive map), linear or fuzzy models.

**Weighted signed digraphs - deterministic cognitive map.** Cognitive map (CM) is a directed graph (digraph) $G(V, E)$ with vertex set $V(G)$ and edge set $E(G)$, which includes two functions $F_V : V(G) \to \mathbf{R}$ and $F_E : E(G) \to \mathbf{R}$. The CM is uniquely defined by the adjacency (or weight) matrix $W = [w_{ij}]_{n \times n}$, where $n$ is the number of vertices and $w_{ij}$ is the weight of the arc $e_{ij} \in E(G)$; $w_{ij} = 0$ means that the corresponding vertices are not connected.

The vertices $C_i \in V(G)$ of CM correspond to concepts (factors) that characterize a situation (or a system), and the arcs define causal connections between concepts. Each concept represents a characteristic of the situation. In general, it can stand for events, actions, goals, values, trends of the situation which is modeled as a cognitive map.

Each vertex-concept is characterized by a number $A_i$ (the function $F_V$) that represents its value and it results from the transformation of the real value of the situation's variable, for which this concept stands.

Between concepts there are three possible types of causal relationships that express the type of influence going from one concept to the others. The weights of the arcs $e_{ij}$ between concept $C_i$ and concept $C_j$ (the function $F_E$) could be positive, $w_{ij} > 0$, which means that an increase in the value of concept $C_i$ leads to the increase of the value of concept $C_j$, and a decrease in the value of concept $C_i$ leads to the decrease of the value of concept $C_j$; or, there could be negative causality, $w_{ij} < 0$, which means that an increase in the value of concept $C_i$ leads to the decrease of the value of concept $C_j$, and vice versa (Fig. (**5**)).

Mathematical model of CM consists of an $1 \times n$ state vector **A** which includes the values of the $n$ concepts and $n \times n$ weight matrix $W$ which gathers the weights $w_{ij}$ of the connections between the $n$ concepts of CM. The matrix $W$ has $n$ rows and $n$ columns where $n$ equals the total number of distinct concepts of CM and the matrix is considered to have diagonal elements equal to zero, since it is assumed that no concept causes itself.

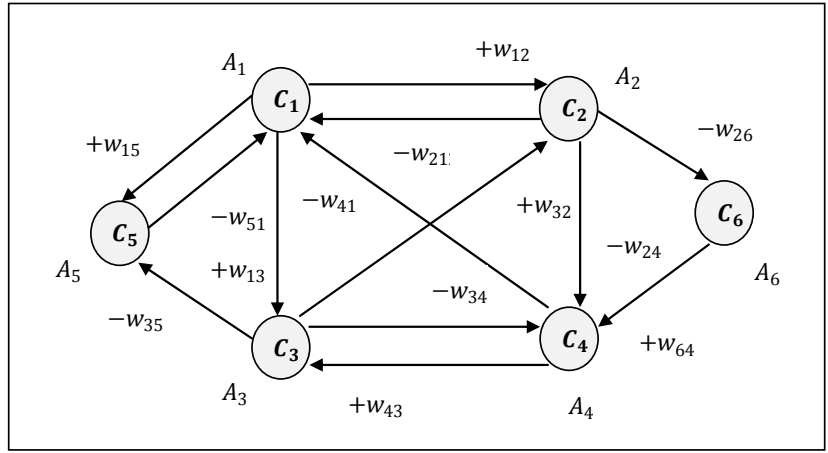

**Figure 5:** A simple model of cognitive map.

The value of each one concept is influenced by the values of the connected concepts with the appropriate weights and by its previous value. So, the value $A_i$ for each concept $C_i$ is updated by the following rule:

$$A_i(t+1) = f\left(\sum_{\substack{j=1 \\ j \neq i}}^{n} A_j(t)\, w_{ji}\right) + A_i(t)$$

where $A_i(t+1)$ is the value of concept $C_i$ at time $t+1$, $A_j$ is the value of concept $C_j$ at time $t$, $A_i(t)$ is the value of concept $C_i$ at time $t$, and $w_{ji}$ is the weight of the connection between $C_j$ and $C_i$, and $f$ is a threshold function.

Thus, in vector form

$$\mathbf{A_{t+1}} = f(\mathbf{A_t} \circ W) + \mathbf{A_t}$$

So, the new state vector $\mathbf{A_{t+1}}$ is computed by multiplying the previous state vector $\mathbf{A_t}$ by the weight matrix $W$. The new vector shows the effect of the change in the value of one concept in the whole CM. The above equality includes the value of each concept at previous time instants, so the CM possesses memory capabilities and there is smooth change after each new updating step of CM.

**Linear models.** The behavior of linear models is associated with the notion of increment value $p_i(t+1) = A_i(t+1) - A_i(t)$ of the factor $C_i$; the increment value in linear models is referred to as an impulse. The increment value is computed by the formula

$$p_j(t+1) = \sum_{i \in I} w_{ij}\, p_i(t)$$

where $I$ is the set of all vertices from which there are paths to the vertex $v_j$.

The most important characteristic of linear model is stability. The vertex $v_i$ is called pulse-resistant if its impulse is bounded, that is, there exists a number $\delta > 0$ such that $|p_j(t+1) < \delta|$ for every $\delta$. The vertex $v_i$ is called absolutely pulse-resistant if its value $A_i(t)$ is bounded. The graph is called (absolutely) pulse-resistant if all its vertices are stable in the appropriate sense. The analysis of stability is conducted in terms of eigenvalues of the incidence matrix of the graph. It is known, that the graph is pulse-unstable if there exists an eigenvalue exceeding $|1|$; the graph is absolutely stable if and only if it is pulse-resistant for every pulse process and there is no eigenvalue equal to 1 among all its eigenvalues.

To stabilize the unstable graphs, different methods can be used. One method is to change the adjacency matrix (adding or removing the arcs) of strongly connected components of the graph; another one is to construct an additional graph-regulator that closes, by means of feedback loops, the inputs and outputs of the initial graph.

The drawback of linear models, essential for applications, consists in rigidity of these models; this means that it is necessary to specify the weights in the form of precise numerical values, whereas the main feature of ill-structured situations is impossibility to obtain the reliable numerical estimates of the weights. So, fuzzy models serve as the most adequate models as alternatives to linear models.

**Fuzzy models - fuzzy cognitive maps.** Cognitive maps were introduced by a political scientist R. Axelrod [76] who used cognitive maps for representing social scientific knowledge and describing the methods that are used for decision making in social and political systems. Then B. Kosko [70, 77] enhanced the power of cognitive maps, considering fuzzy values for the concepts of cognitive map and fuzzy degrees of interrelationships between concepts. After this pioneering work, fuzzy cognitive maps attracted the attention of scientists in many fields and have been used in a variety of different scientific problems. Fuzzy cognitive maps have been used for planning and making decisions in the field of international relations and political developments [78] and for analyzing graph theoretic behavior [79], been proposed as a generic system for decision analysis [80] and for distributed cooperative agents [81]. Fuzzy cognitive maps have also been used to analyze electrical circuits [82], to structure virtual worlds [83]. In the control related themes fuzzy cognitive maps have been used to model and support plant and manufacturing control [84, 85], to represent failure models and effects analysis for a system model [86, 87] and to model the supervisor of control systems [88], to model dynamic system control [89, 90]. It is obvious that there is high interest in the use of fuzzy models in a wide range of different fields.

As with the deterministic cognitive maps, fuzzy cognitive map theory makes use of a symbolic representation for the description and modeling of system. It utilizes concepts to illustrate different aspects in the behavior of the system and these concepts interact with each other showing the dynamics of the system. A fuzzy cognitive map integrates the accumulated experience and knowledge on the operation of the system, as a result of the method by which it is constructed, *i.e.*, using human experts' knowledge related to the operation of system and its behavior in different circumstances [91, 92].

It should be mentioned that since all the values in the digraph are fuzzy, so concepts take values in the range between (0, 1) and the weights of the arcs are in the interval $(-1, 1)$, or, in general, are the values from a linguistic scale (linearly ordered set of linguistic values that describe all possible strengths of influences).

Let $C_{j1}, C_{j2}, \ldots, C_{jk}$ be the set of all factors which are the inputs for the factor $C_i$, *i.e.*, they are the vertices of the arcs entering $C_i$. Then, in general case, the value of $A_i$ at time $t+1$ depends on the values of input factors at time $t$ and on the weights of the arcs that connect these factors with the factor $C_i$:

$$A_i(t+1) = f_i\big(A_{j1}(t), A_{j2}(t), \ldots, A_{jk}(t);\ w_{j1,i}, w_{j2,i}, \ldots, w_{jk,i}\big)$$

The selection of the functions $f_i$ (influence functions, aggregation functions), which in general case can be different for different vertices, defines every concrete model. Usually, in applications the simpler models are considered, when these functions are the same for all factors. Such models are called homogeneous and to solve problems of analysis of situations within the framework of these models, matrix methods can be applied.

The process of cognitive maps building precedes the work with them. Obviously, this process cannot be completely formalized and to large extent is subjective. On the other hand, it can be seen as a process of extracting knowledge from experts, and results are presented in the form of cognitive maps. The components of this process are the construction of the conceptual structure of the problem domain (identification of key factors and relationships between them), the selection of aggregation function, and the construction of linguistic scales (for fuzzy models) and the assignment of weights to the arcs.

Cognitive maps as models of representation of expert knowledge reflect a subjective vision of a situation. Different experts can build different cognitive maps for the same problem domain; these maps can differ not only by the values and signs of weights, but also by the set of factors. In this sense, cognitive maps can be considered as a tool of rough, qualitative description of situation. Cognitive models and cognitive analysis with some confidence may indicate a possible trend of the situation as a result of various control actions, identify the various side effects of solutions, but in principle cannot give guaranteed estimates and predictions. The adequacy of a cognitive model can be judged only by the results of its application.

Let us note some actual directions for further development of cognitive approach:

- Modeling the development and control of development of dynamic situations under resource-limited settings;

- Modeling of conflict situations, threats and counter threats in terms of cognitive maps;

- Structural analysis of cognitive maps: identification of unwanted cycles, stability analysis (*i.e.*, robustness - insensitivity to small perturbations) of a given situation, *etc.*;

- Physical time in cognitive models;

- Heterogeneous cognitive maps with different influence functions for different factors, and methods of their analysis;

- Building techniques for cognitive maps with help of typical structures and procedures;

- Studying the prediction reliability in cognitive models;

- Studying the adequacy of various models of cognitive maps in different problem domains;

- Building of space-time cognitive models, multidimensional and multilevel hierarchical cognitive maps and methods of their analysis.

It should be noted that the theory of (fuzzy) cognitive maps is a powerful tool for modeling, analysis and control in weakly-formalized and ill-structured problem domains, as well as in domains with uncertainties and information incompleteness.

The method of cognitive modeling is the method of soft modeling and simulation. The closest analogues of this method are simulation modeling and the method of system dynamics. The advantage of the method of cognitive modeling is that it can operate not only with precise quantitative values and formulas, but with the qualitative values and estimates as well. But this advantage can also be in some sense regarded as a drawback, since the results are qualitative.

Cognitive models help quickly get the initial results, more detailed understanding of simulated system, and identify regularities and then move on to more accurate models (if this is possible and necessary). Therefore, the most reasonable and efficient level of application is the use of cognitive modeling at the top level of decision-making, upon the analysis of complex socio-economic, technical, techno-economic, and other weakly-formalized and ill-structured systems.

### 4.7. Multi-Agent Models

Discrete individual-based or agent-based modeling [93-95] has become a very promising and powerful methodology to describe the occurrence of complex behavior in diverse systems. The advantage of such an individual-based approach is given by the fact that it is applicable also in cases where only a small number of agents govern the system evolution.

Agent-based model is the real world construed from many separate active subsystems, or units. Each unit interacts with other units, which constitute for the unit the external environment, and this unit, in the process of operation, may change both the external environment and its behavior.

In agent-based modeling, a system is modeled as a collection of autonomous decision-making entities called agents. Each agent individually assesses its situation and makes decisions on the basis of a set of rules. Agents may execute various behaviors appropriate for the system they represent, for example, producing, consuming, or selling. Repetitive competitive interaction mode between agents is the main feature of agent-based modeling, which relies on the power of computers to explore dynamics out of the reach of pure mathematical methods [96, 97]. At the simplest level, an agent-based model consists of a system of agents and the relationships between them. Even a simple agent-based model can exhibit complex behavior patterns and provide valuable information about the dynamics of the real-world system that it emulates. In addition, agents may be capable of evolving, allowing unanticipated behaviors to emerge. Sophisticated agent-based models sometimes incorporate neural networks, evolutionary algorithms, or other learning techniques to allow realistic learning and adaptation.

There are many definitions of an agent. Common in all these definitions is that the agent is an entity that is active, with autonomous behavior, can make decisions in accordance with a certain set of rules, can interact with the environment and other agents, and can evolve [95, 98, 99].

The purpose of agent-based models is to get an overview of these global rules, the general behavior of the system, based on assumptions about the individual, the local behavior of its individual active sites and the interaction of these objects in the system.

Multi-Agent System (MAS) may be considered as an intelligent tool for the solution of such problems as planning, scheduling, decision making and control in the framework of production processes [105, 106]. The MAS approach seems to be the most feasible. It respects the complicated characteristics of the goal that one aims to achieve. There are some significant reasons that can motivate one to choose the MAS approach to the solution of complex problems and decision making, such as:

- *Modularity*. Each agent is an autonomous module and can work without interventions of the external world. Each agent can possess different capabilities or functionalities and through cooperation the agents are able to achieve a variety of goals.

- *Parallelism*. The MAS approach enables to work in parallel. A complicated problem could be solved in an acceptable time by using a number of agents, *e.g.*, gathering information from various resources allocated in different places.

- *Flexibility*. The MAS approach is able to react in a flexible manner to each change in the environment. Through cooperation the agents can assist each other to compensate the lack of capability or knowledge. They can share information or own capacity to resolve a newly appeared situation, if one agent is not able to do so. Besides that, each intelligent agent can do reasoning about whom and when it has to cooperate with, in order to achieve the effective performance.

It is also supposed that agents have the following properties:

- *Autonomy*. Each agent, as mentioned before, thinks and acts locally. It means that agent operates without direct interventions from other agents to achieve its own goals.

- *Social ability*. Agents can cooperate with other agents to achieve common goals.

- *Reactivity*. Agents react on changes in environment; it is needed to describe the negotiation process.

- *Pro-activeness*. Agents do more than response on events generated by environment; they can show goal-directed behavior.

Generally, multi-agent systems consist of the following main components:

- A set of system units, which is divided into two subsets, active and passive; members of the active subset, active units, called agents that manipulate the members from the passive subset, passive units called objects;

- Environment, or space, in which agents and objects have to function and act;

- Tasks (functions, roles), which are assigned to agents;

- Relations, interactions, between agents;

- Organizational structures, configurations, which are generated by the agents;

- Agent actions, for example, different operations over objects, communication acts.

Interaction between agents means establishing bilateral and multilateral dynamic relations between the agents. Interactions between the agents have a definite direction - positive or negative, *i.e.*, they have the character of assistance or resistance, attraction or repulsion, cooperation or competition, cooperation or conflict, coordination or subordination, *etc.* Interconnections and interdependencies between agents are characterized by certain intensity. Interactions between agents are dynamic. Multi-agent models are used to study decentralized systems, the dynamics of functioning of which is determined not by global rules and laws, but on the contrary, these global rules and laws are resulted from individual activity of a group of agents. Typically in such systems there is no global centralized management, agents operate under their own laws asynchronously.

Agent-based models capture emergent phenomena. Emergent phenomena result from the interactions of individual entities. By definition, they cannot be reduced to the system's parts: the whole is more than the sum of its parts because of the interactions between the parts. An emergent phenomenon can have properties that are decoupled from the properties of the part.

Highly flexible nature of agent-based technique allows one to apply it in many cases, especially when there is potential for emergent phenomena:

- Individual behavior is nonlinear and can be characterized by thresholds, if-then rules, or nonlinear coupling. Describing discontinuity in individual behavior is difficult with differential equations;

- Individual behavior exhibits memory, path-dependence, and hysteresis, non-markovian behavior, or temporal correlations, including learning and adaptation;

- Agent interactions are heterogeneous and can generate network effects. Aggregate flow equations usually assume global homogeneous mixing, but the topology of the interaction network can lead to significant deviations from predicted aggregate behavior;

- Averages do not work. Aggregate differential equations tend to smooth out fluctuations, not agent-based model, which is important because under certain conditions, fluctuations can be amplified: the system is linearly stable but unstable to larger perturbations.

Or when describing the system from the perspective of its constituent units, *i.e.*, when

- The behavior of individuals cannot be clearly defined through aggregate transition rates;

- Individual behavior is complex. Everything can be done with equations, in principle, but the complexity of differential equations increases exponentially as the complexity of behavior increases. Describing complex individual behavior with equations becomes intractable;

- Activities are a more natural way of describing the system than processes;

- Validation and calibration of the model through expert judgment is crucial. Agent-based model is often the most appropriate way of describing what is actually happening in the real world, and the experts can easily "connect" to the model, modify it and make improvements;

- Stochasticity applies to the agents' behavior. With agent-based model, sources of randomness are applied to the right places as opposed to a noise term added more or less arbitrarily to an aggregate equation.

Moreover, the flexibility of agent-based models can be observed through many other aspects. For example, it is easy to add more agents to an agent-based model or delete agents. Agent-based models also provide a natural framework for adjusting the complexity of the agents: behavior, degree of rationality, ability to learn and evolve, and rules of interactions. Another aspect is the ability to change levels of description and aggregation: one can easily play with aggregate agents, subgroups of agents, and single agents, with different levels of description coexisting in a given model. One may want to use agent-based modeling when the appropriate level of description or complexity is not known ahead of time and finding it requires some tinkering.

Multi-agent approach allows one to investigate the problem of collective interaction, effectively solve prediction problems, investigate the processes of self-organization, and also allows natural descriptions of complex systems [100-107].

So, the agent-based modeling is a powerful simulation modeling technique for studying the behavior and evolution of complex systems. Presently, the use of the multi-agent systems paradigm for the modeling of complex systems has many successful applications, as it allows specialists to gather information quickly and process it in various ways.

## 5. HIERARCHICAL STATE GRAPH DIAGRAMS

The above review reveals that conventional methods in modeling and control of systems have contributed a lot in the research and the solution of many complex problems, but their contribution to the solution of the increasingly complex dynamical systems in ever changing environments will be limited. It has become quite clear that the requirements in the control and modeling of systems cannot be met with the existing conventional control theory techniques and it is necessary to use new methods that will exploit past experience, rely heavily on expert knowledge-bases, will have learning capabilities and will be supplied with intelligent simulation tools [108-119].

The main idea behind attempts to cope appropriately with complex problems is to describe relationships between various elements (objects, agents, parameters, variables, *etc.*) of a system and thus provide knowledge representation and inference. Such description of a problem should utilize experts' beliefs and cognition about a problem, yielding thorough analysis, reliable forecasting and decision-making. Problem solving planning and control in complex systems is an activity which deals with problems related to (optimal) selection of control actions, operations, system parameters, *etc.*, required for achievement of prescribed goals. Optimality of generated solution depends significantly upon the problem domain expert experience, beliefs, and knowledge about the system behavior [114, 115].

In addition, in real world (complex) situations we face two important issues. On one side, we find causal and structural dependencies between system variables; on the other side, different experts provide different views of the same problem, they build different models of situation development and thus propose different solutions, which in turn may perform different output and control effects. Modeling taxonomy based on graphical representation of situation (object, system, *etc.*) development and structural relations in concrete problem domain by means of state diagrams [121, 123] allows for taking into account of each possible view of a problem and thus modeling and then generate global model by augmenting a set of separate state diagrams [122, 123].

Providing a description of each object state development and then the overall complex system behavior, based on knowledge bases and domain experts' experience and monitoring data, state transition diagrams enable thorough analysis and come up with an answer to "what-if" question. An "if" input vector is composed of a set of control actions on lower level objects; however, our model, Hierarchical State Graph Diagrams - HSGD [120-123], does not restrict us in selection of control objects and the input vector components can have a distributed nature and be applied at different levels of abstractions. What happens when an input vector affect a system, we can find as a hierarchical state diagram (hierarchical network) output, suggesting possible actions that steer a system to the desired states or bring the system to equilibrium or stable states. That is, state diagrams technique is mostly qualitative tool [124] but also provides some quantitative analysis of system development, but it cannot present exact mathematical answer but rather to point out the gross behavior of a system, to show the global patterns of how the whole of our hierarchical network of state diagrams behaves, and to get a view of what goals can be achieved by means of applying one or another set of control actions.

The methodology that we propose aims to contribute to the overcoming of shortcomings of approaches based on traditional knowledge-based systems that lead to expert system failure related to, for instance, incomplete, inconsistent, and ambiguous knowledge bases. Another shortcoming that troubles conventional approaches to complex systems modeling is related to the type of relationships between system parameters and variables. These appear quite often to be rather causal and structural than just the only explicit "if-then" rules. Moreover, depending on a system we are modeling, its dynamics can additionally bring difficulties in expert knowledge base development.

The aim of the approach is to promote the methodology for dynamic modeling, multi-criteria decision analysis and reasoning about potential effects of initially generated system development plans and related parameters with respect to control strategies. One technique for modeling systems which contributes to the effort for more intelligent simulation and control methods is based on the HSGD technique.

**5.1. Conceptual Basics of Agent Oriented Hierarchical Parametric Modeling**

We use several concepts [121, 122, 125, 126] to model complex systems. They simplify complex problem solving in multi-parameter setting and divide it into more simple, manageable and understandable units.

The modeling paradigms, which are the most close to our approach, use the concepts of modeling complex systems in an agent- and object oriented point of view, and discrete event and finite state machines methodologies. The concepts are as follows:

- *Decomposition*: divides a large complex system consisting of a set of interrelated subsystems (objects, elements, agents) into set of small subsystems. The small subsystems can be easily modeled, executed and are manageable. It also divides the problem environment into a set of separate problem domains. By integrating (models of) small subsystems and considering them in each of the problem domains, we can build the large system for the whole problem environment. In the agent oriented view, we divide the complex system into a collection of sets of agents, which are studied in every specific problem domain (Fig. (**6**)).

- *Abstraction*: represents complex system into simple systems conceptually. It focuses on the interested parts of complex system. Other detail parts are hided for simplicity. Abstract representation provides for qualitative description and analysis of complex system. Using conceptual models that have only essential features necessary to represent or solve complex problem, we can understand and analyze complex system.

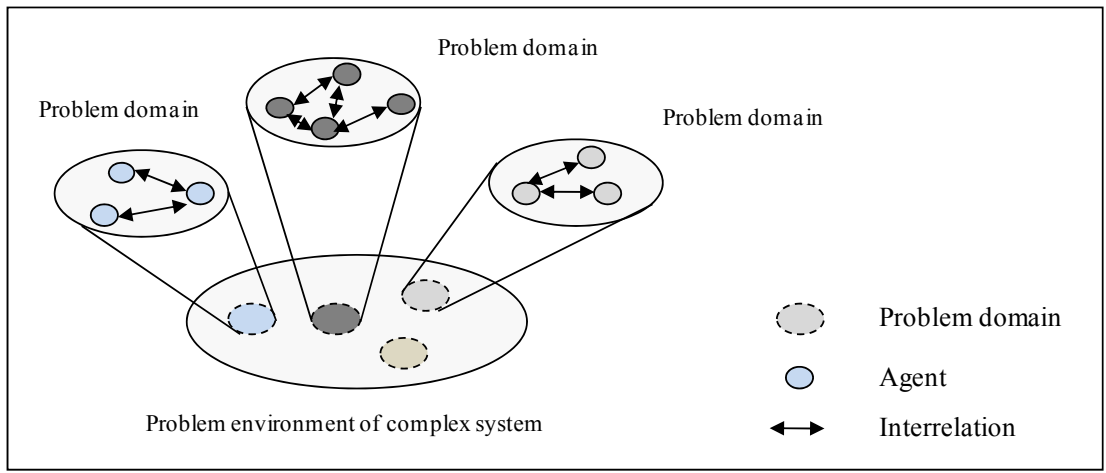

**Figure 6:** Problem environment and its representation in complex system.

- *Polymorphism*: introduces a set of polymorphic parameters which are applied to a collection of system agents. It enables one to turn from the concept of modeling and control at object level to the concept of modeling and control at the level of classes of objects. This helps turn form individual models to the integral models at arbitrary level of abstraction.

- *Hierarchy*: represents the system agents, the problem and control domains and their qualitative characteristics, *i.e.*, polymorphic parameters (Fig. (**7**), (**8**)). The principle of hierarchy allows us to structure agents relations in a system and distinguish the essential interrelations in the system for aggregation and scaling (recount) of dependent parameters, and also helps structure the problem and control domain of an object.

- *Layering*: separates system functions into layers according to their roles or responsibility. Complex system can be represented as a layered system in which each layer plays its role. In general, layering concept helps understand and design complex system.

The above concepts lead to the model s*calability* [120, 123] that provides simultaneous representation of goals, parameters and development plans of various objects, which is very important in the problem of building the models of development of large complex systems (Fig. (**8**)).

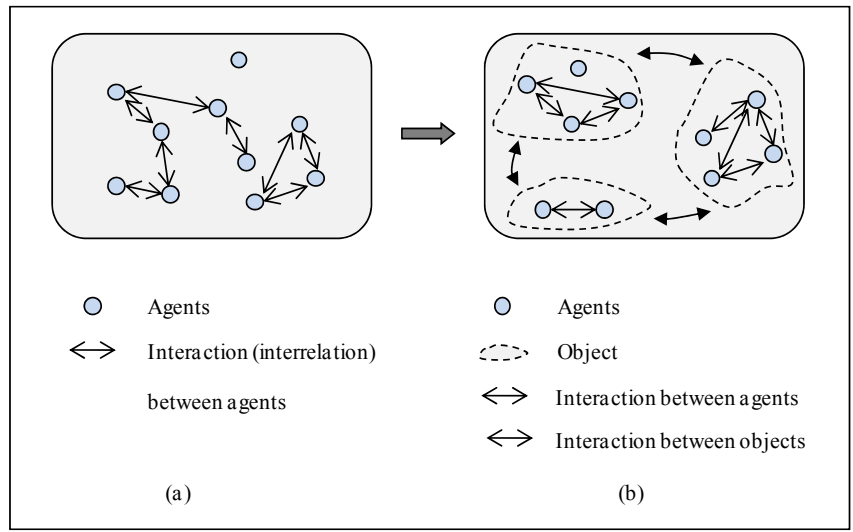

**Figure 7:** Principle of hierarchization: (a) flat network structure, (b) hierarchical structure.

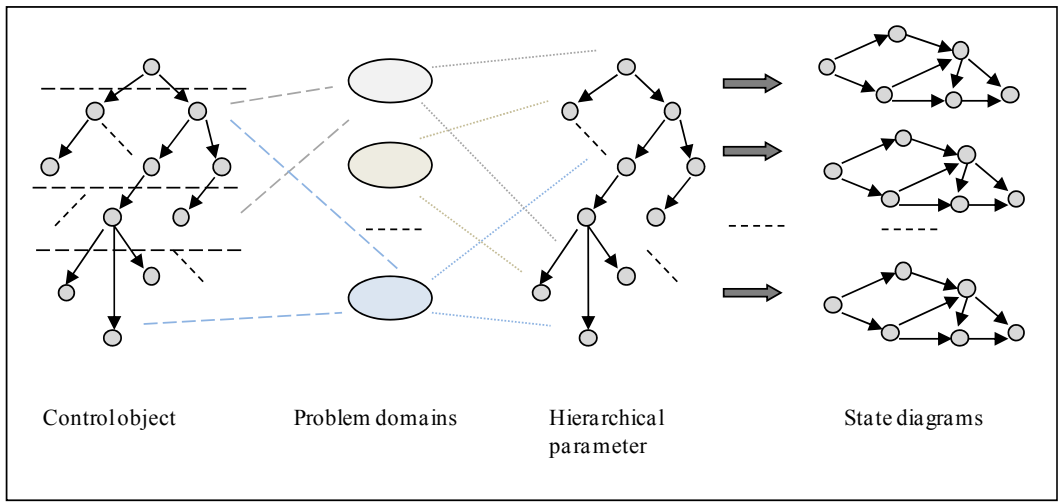

**Figure 8:** Generalized structure of the hierarchical decomposition of development model.

The important features of our model are formulated as follows:

- Model of development of each object at any level of hierarchy is sufficiently autonomous. This provides a sufficient degree of decomposability and therefore results in flexibility of building large models;

- Objects in a system that we model, being separate units, are not modeled as such separate units but as classes, within each class the objects have common development goals;

- Models of development of objects at different levels are information compatible, that is, outputs of one object model at some level serve as the input for another object model at some other levels.

The principle of hierarchy is common in the real-world systems. There are enormous number of examples of it, beginning from ecosystems and biosystems to neuronal ensembles and cognitive networks to organizational and political systems to social and economic ones. So, the emergent (collective) characteristics of a particular lower level system frequently form an individual object (agent) at a higher level of the hierarchical system. This aspect has been emphasized by many researchers on artificial intelligence and complex systems [127].

Our model HSGD accommodates a hierarchical multilevel multi-agent dynamical system, in which an agent can itself be a collection of other agents. The HSGD structure is as follows:

- HSGD is composed of a number of levels. Each level consists of a number of dynamical multi-agent systems which describe the behavior composed of sets of agents;

- Each set of $(n-1)$ level agents is aggregated into a $n$ level agent;

- Behavior and goal achievement of $n$ level agent emerges from the organization of inter-level dependencies and the behavior and goal achievements of 1 to $n$ level agents.

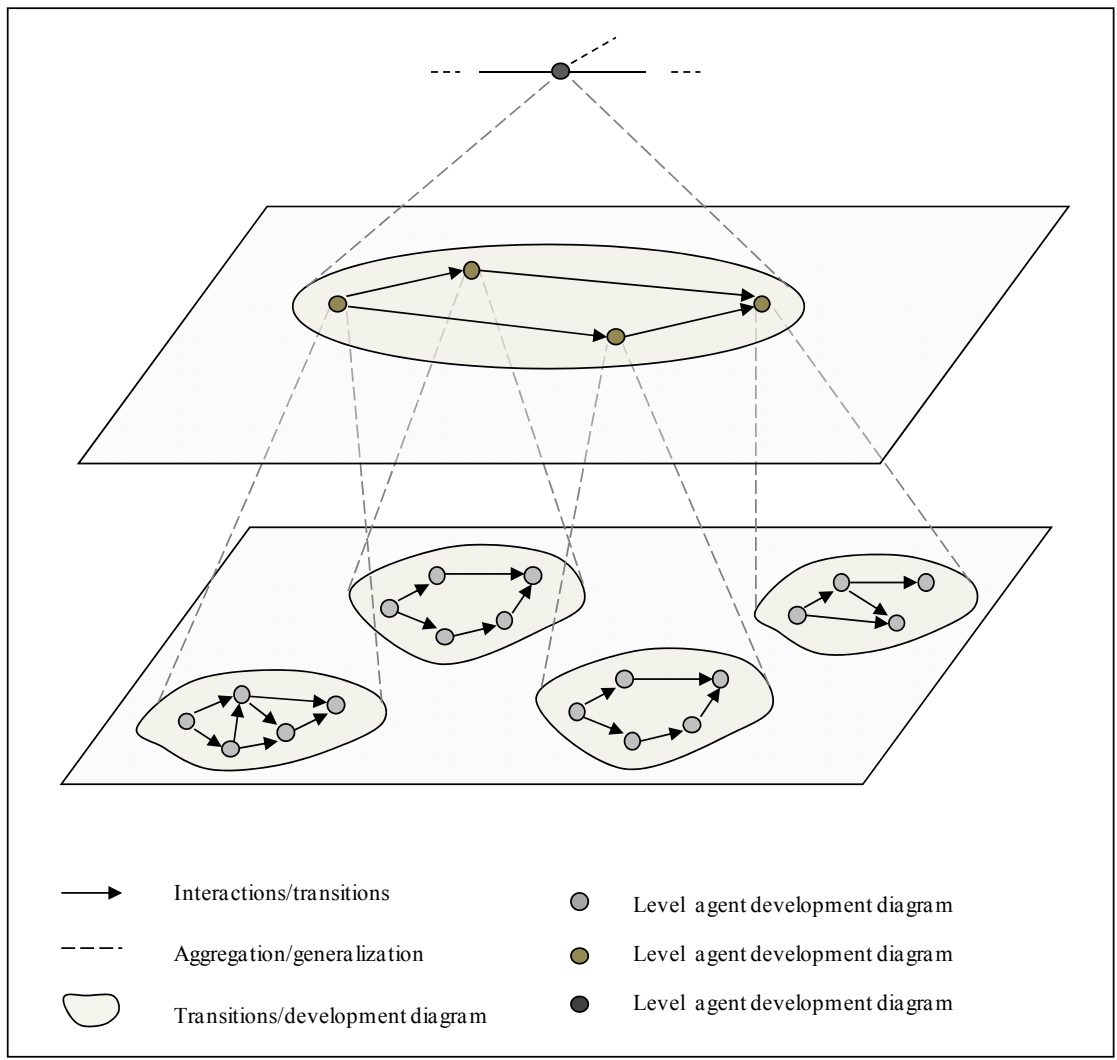

**Figure 9:** Hierarchical state graph diagrams model using multi-agent multilevel dynamical system in nested hierarchies.

HSGD naturally admits a description in terms of higher level and lower level, where the lower level is nested within the higher level. Any agent at any level is both a component of a given collection in its own level and a subsystem decomposable into a set of other agents at its adjacent lower level of HSGD (Fig.

(**9**)). Note that the agents in the lowest level are the minimum units, which are indecomposable of this hierarchical system. HSGD is a heterogeneous system where each set of agents in each level is evolved in its own goal direction and adapts to the requirements of problem domain environment through the application of its own development plans. The interaction topology of HSGD can also be heterogeneous hierarchical structures. Namely, the development rules, the inter-level dependencies rules and the interaction topology of distinct sets of agents can be different, and these different schemes hierarchically construct the HSGD model and lead to the hierarchical network of dynamical processes which, properly defined, leads to intelligent description of behavior. In mathematical terms, the HSGD model can be in generalized form defined as follows:

$$\mathbf{HSGD} = \{A, D, R, L, T, O, C\}$$

where

$A$: agents grouped in objects that lie at various hierarchical levels;

$D$: dynamic systems represented as state graph diagrams that model development plans for each group of agents (object);

$R$: coordination rules for dynamic systems at different (adjacent) hierarchical levels;

$L$: the number of levels in hierarchy;

$T$: the hierarchical interaction topology of HSGD given by the rules of inter-level dependencies;

$O$: the objects of each class in each hierarchical level;

$C$: the control scenarios/strategies used for each class of objects to achieve their goals.

Considering the problem of model building and simulation of complex system, we distinguish global goals lying at higher levels of hierarchy and local goals lying at the lower levels. The goals of objects at different levels of hierarchy are highly interconnected. Interaction between higher-level object and lower-level objects is such that the achievement of goals of lower-level objects immediately influences the achievement of goal of the higher level object. This leads to the concept of *hierarchical state* which corresponds to the global goal of higher level (Fig. (**10**)). This means that each of the objects or a class of objects is immersed onto the intersection of states that correspond to the concrete set of the objects' goals, but the whole system is immersed into the hierarchical state. This approach is applicable at any level of hierarchy. The approach is useful for different kinds of systems [128, 129].

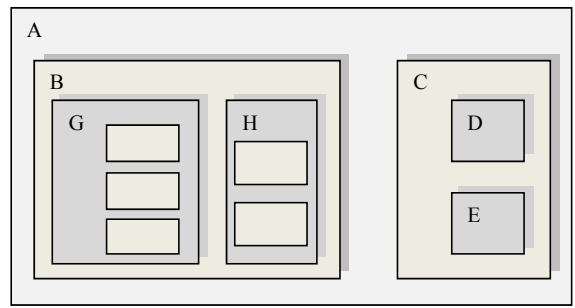

**Figure 10:** The scheme of decomposition of a hierarchical state.

Using the appropriate semantic interpretation, the hierarchical state shows how the current states of objects of different levels of hierarchy are related to each other. Our approach to the representation of states has the following useful properties:

- Aggregation of parameters and therefore simplification of modeling of complex system;

- Formalization of information gathering and its assessment for getting integrated and individual estimates and global and local evaluations of the system at any level;

- Ability to analyze the multi- and inter-level dynamics of system and study different aspects of inter-connected dynamic processes in a unified way.

**5.2. Model of State Diagrams**

The main idea underlying the abstract representation of control and development processes in multilevel dynamic system relies on the use of the model of state diagrams [125]. State diagrams provide the qualitative description of dynamics of parameters and of controlled state dynamics of objects with use of control scenarios. The use of state diagrams technique supposes the initial classification of control objects over the system state space and the construction of canonical model of state dynamics represented by the model of state transition diagram. The state space is defined as follows (Fig. (**11**)). The continuous time interval is divided into parts by means of the interval partition. On each time interval we have some subsets of states, in which objects remain during this interval. Every state is defined by the vector of values of corresponding parameters. The state of the object is thought of as a situation which is described by the state vector of each agents of the object. The canonical model is constructed by using (1) databases containing the information about the dynamics of values of parameters that characterize the set of objects at a given time interval, and (2) expert knowledge-base consisting of declarative and procedural knowledge containing the rules of classification.

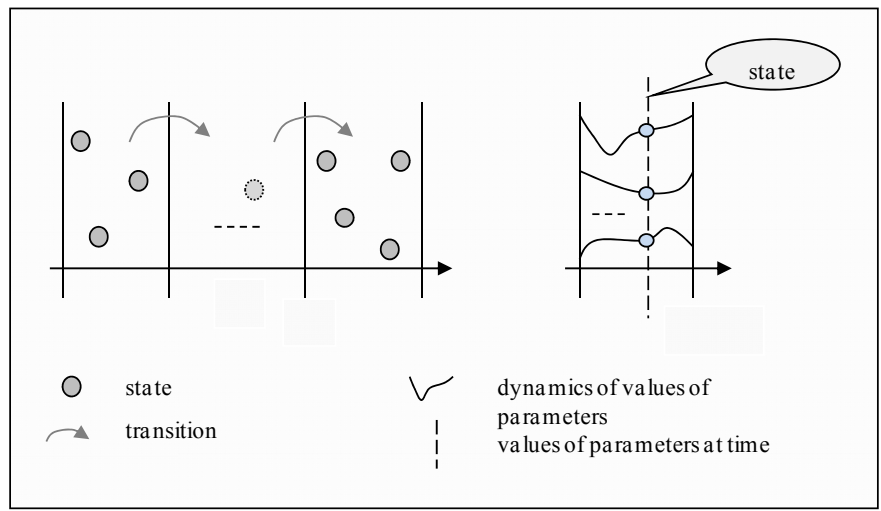

**Figure 11:** The definition of state of an object.

The state transition diagram (Fig. (**12**)), in terms of which the canonical model is described, is defined as:

$$D = (\{\Delta, \pi\}, \{S, S_0, S^*\}, K, P^+, \{\mu_i, \mu_0, \mu^*, i = \overline{1,n}, n = card(S)\})$$

where

$\{\Delta, \pi\}$: the finite time interval $\Delta$ divided into parts $\Delta_j$ by the partition $\pi = (\tau_0, \tau_1, \tau_2, \ldots, \tau_n)$, where $\Delta_j = \tau_j - \tau_{j-1}$ and $\bigcup_{j=1}^{n} \Delta_j = \Delta$, $\bigcap_{j=1}^{n} \Delta_j = \emptyset$;

$\{S, S_0, S^*\}$: the set of states ordered by the classification rule $K$, $S = \{S_j, j = \overline{1,n}\}$, $S_j \in S^{\Delta_j}$; more precisely, the rule $K$ divides $S$ into $n$ equivalence classes such that we get $S = \{S_{jl}: j = \overline{1,n}; l = card(S^{\Delta_j})\}$, $card(S) = m = l \times n$, $S_{jl} \in S^{\Delta_j}$, and $\bigcup_{j=1}^{n} S^{\Delta_j} = S$, $\bigcap_{j=1}^{n} S^{\Delta_j} = \emptyset$;

$S_0, S^*$: the initial and final state respectively;

$P^+$: the set of weighted arcs that determine state transitions, each arc is assigned the transition time $\theta$; if an arc $e = (S_i, S_j) \in P^+$ then $S_i \prec S_j$, where $\prec$ is the order relation determined by $K$, i.e. there is an arc $e = e_{ij}$ from $S_i$ (previous state) to $S_j$ (subsequent state);

$\{\mu_i, \mu_0, \mu^*\}$: the distributions $\mu_1, \mu_2, \ldots, \mu_n$ over the vertices-states of the diagram at time moments $t_i \in \Delta_i$; without loss of generality, one can take $t_i = \tau_i$;

$\mu_0$: is the initial distribution;

$\mu^*$: is the final distribution.

The canonical model is used as a tool for formalization of qualitative properties of multilevel dynamical system and represents hypothetical model, based on expert knowledge, of state dynamics of a set of objects. The canonical model describes the qualitative character of state dynamics, using the distributions $\mu_i$ of the objects; the distributions determine the states in which the objects at some time $\tau_i$ can be found.

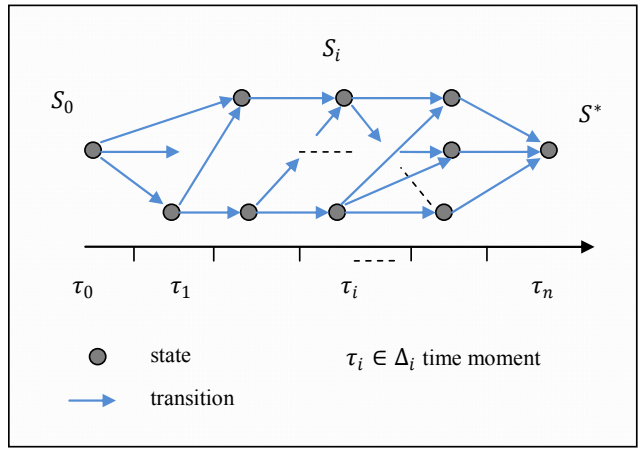

**Figure 12:** The diagram of canonical model of state dynamics.

The canonical model is then used for comparative analysis with the actual state dynamics of the set of objects. For this purpose, the states of the objects are consecutively re-estimated at time moments $\tau_i$ in order to get the actual distributions of the objects over the states of canonical model, and then to compare this distribution with the required one. This helps represent the core essence of system control problems and state dynamics of control objects. The description of actual process of state dynamics at arbitrary time interval $\Delta_j$ is based on the use of states of the canonical model as objects' classifier.

The state transition diagram of actual state dynamics (Fig. (**13**)) is represented as follows:

$$\boldsymbol{D_a} = (\{\boldsymbol{D}\}, \boldsymbol{P^-}, \boldsymbol{\eta(t)}, \boldsymbol{N(t)})$$

where

$\{D\}$: the state transition diagrams, each one extended by the components:

$P^-$: the set of arcs, that describe the state backstep. Thus, the set $P$ of $D_a$ is defined as $P = P^+ \cup P^-$, $P^+ \cap P^- = \emptyset$, where $P^+$ is the set of arcs that describe state transitions; if $e_{ij} = (S_i, S_j) \in P^+$ then $(S_i \prec S_j)$, but if $e = (S_i, S_j) \in P^-$ then while $(S_i \prec S_j)$ there is an arc $e = e_{ji}$ from $S_j$ (subsequent state) to $S_i$ (previous state) or $j = i$;

$\eta(t) = (\eta^1(t), \eta^2(t), \ldots, \eta^\nu(t))$: the vector-function called the counter of objects, $\nu = card(P)$; each component $\eta^k(t)$ of the vector determines the number of objects that change their state from $S_i$ to $S_j$, $k \coloneqq (S_i, S_j)$, at some time interval $\Delta_j$; the counters $\eta^k(t)$ are assigned to each arc $e_{ij} = (S_i, S_j) \in P$; The counters assigned to the arcs $P^+$ characterize the intensity of objects' state transitions (positive processes); the counters assigned to the arcs $P^-$ estimate the intensity of objects' state backsteps (negative processes);

$N(t) = (N^1(t), N^2(t), \ldots, N^n(t))$: the vector function, each component of which $N^i(t)$ defines the number of objects having a fixed state $S_i$, $n = card(S)$.

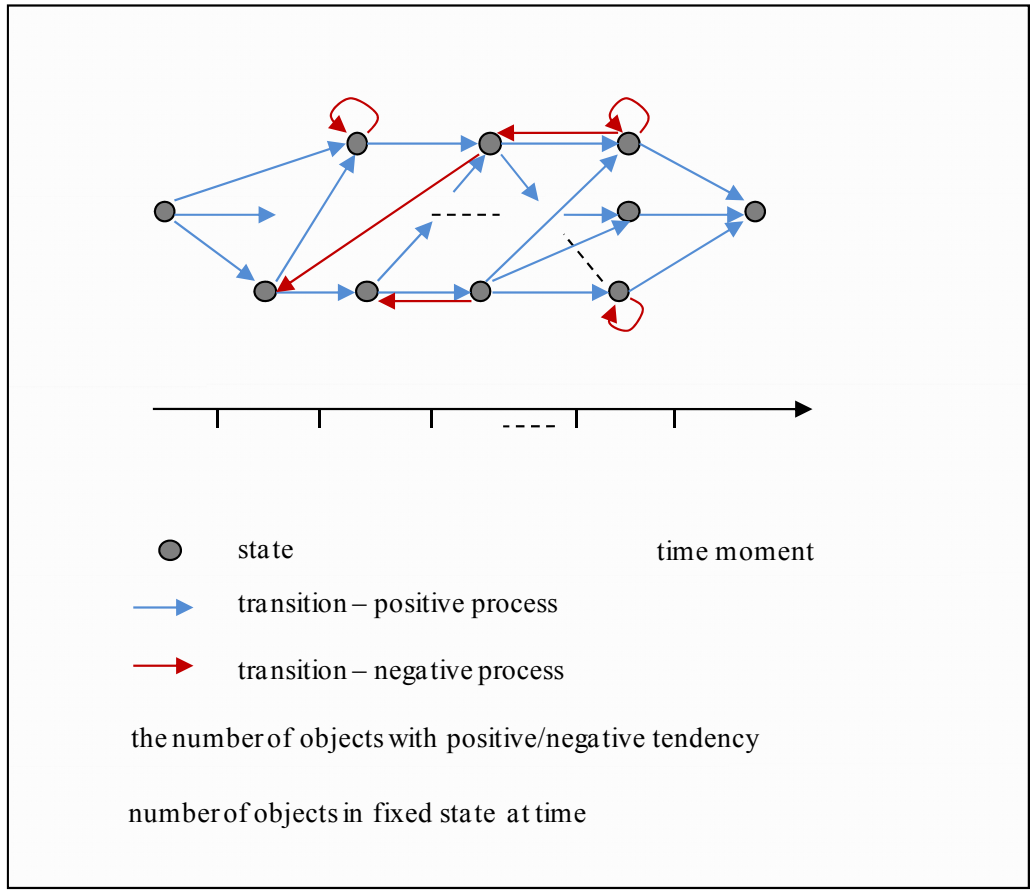

**Figure 13:** The diagram of actual state dynamics.

The functions $\eta(t)$ and $N(t)$ allow us to obtain the information on the relation between processes of development and degradation, and to qualitatively estimate control actions and their efficiency.

**5.3. Coordination of State Diagrams and Their Compositions**

Coordination [130-143] is one of the key issues in modeling, control and design of complex systems, and has been the subject of numerous investigations in areas such as sociology, economics and organizational theory. From an engineering point of view, coordination is conceived as a means to integrate various activities or processes in such a way that the resulting ensemble shows desired characteristics and functionalities. The design of coordination mechanisms is particularly challenging in the field of multi-agent systems, as they are usually embedded in highly dynamic environments, and neither the number nor the behavior of agents can be directly controlled at design time. We argue that additional high-level abstractions need to be integrated into agent oriented design methodologies in order to exploit the full potential of coordination schemes, and to engineer coordinated multi-agent system in open environments in an efficient manner.

Maybe the most widely accepted conceptualization of coordination in the multi-agent field originates from work in the area of organizational science [144] that defines coordination as the management of dependencies between organizational activities. It is straightforward to generalize this approach to coordination problems in multi-agent systems. The subjects whose activities need to be coordinated are the processes in agents. However, the entities between which dependencies arise (or object of coordination) and should be coordinated are usually come down to goals, actions and plans. Depending on the characteristics of the system environment, taxonomy of dependencies can be established, and a set of potential coordination rules are assigned [145]. Within this model, the process of coordination is to accomplish two major tasks: establishing the dependencies in the system, and then, decision respecting which coordination rules must be applied. The coordination mechanism is the way to perform these tasks [146].

We put the problem of coordination within the context of state transition diagrams that model the state dynamics of agents.

We introduce several compositional operations giving the rules of consistency of different state diagrams. These rules help one to construct complex models of state dynamics that combine the interrelations between different sets of parameters and represent the conditions for coordination of state dynamics for objects at different levels of hierarchy. Structural composition of state diagrams provides a synthesis of complex requirements set to the dynamical characteristics of controllable object. The structural composition of simple models is one of the key issues in building the complex models of hierarchical state dynamics.

So, we say that for state diagrams the property of consistency holds if the attainability of certain states takes place in prescribed time-event sequence. The operations of sequential and parallel compositions and operation of aggregation are introduced.

Let $D_1$ and $D_2$ be two state diagrams given at time intervals $\Delta_1 = [0, \alpha_1]$ and $\Delta_2 = [0, \alpha_2]$ correspondingly. Then the operations of compositions for the diagrams are as follows:

a) Sequential composition

   The diagrams $D_1$ and $D_2$ are sequentially composed, that is, they compose a linear fragment if for their time intervals one has $\alpha_1 < \alpha_2$ (Fig. (**14**)).

b) Parallel composition

   The diagrams $D_1$ and $D_2$ are composed in parallel, that is, they compose a parallel fragment if they are defined on the same time interval, $\alpha_1 = \alpha_2$ (Fig. (**14**)).

c) Aggregation

   The aggregation is defined for state diagrams at different levels of hierarchy. In this case, for coordination of state diagrams at neighbor levels of hierarchy the Cartesian product of states of diagrams of lower level of hierarchy is defined (Fig. (**15**)). To do this, one should specify the ordering relation on the subsets of Cartesian product of states of diagrams of lower level of hierarchy.

The above defined compositions are straightforwardly generalized for arbitrary number of state diagrams.

Combining the above two compositions, we get a sequential-parallel composition when the time intervals at which the state diagrams are given has an intersection.

The operation of aggregation can be defined not only on the single states, but also on the subgraphs of state diagrams of lower level of hierarchy.

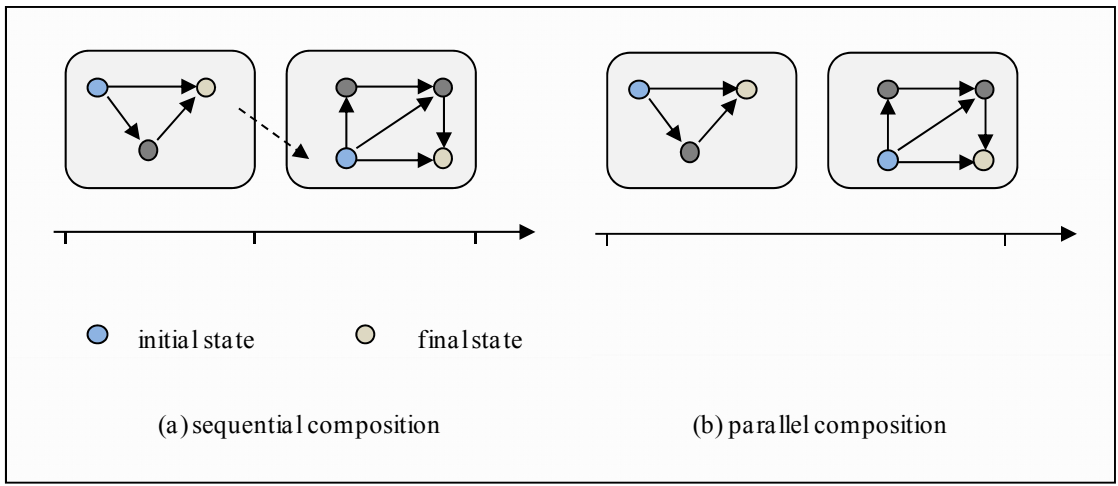

**Figure 14:** Sequential (a) and parallel (b) compositions of state diagrams.

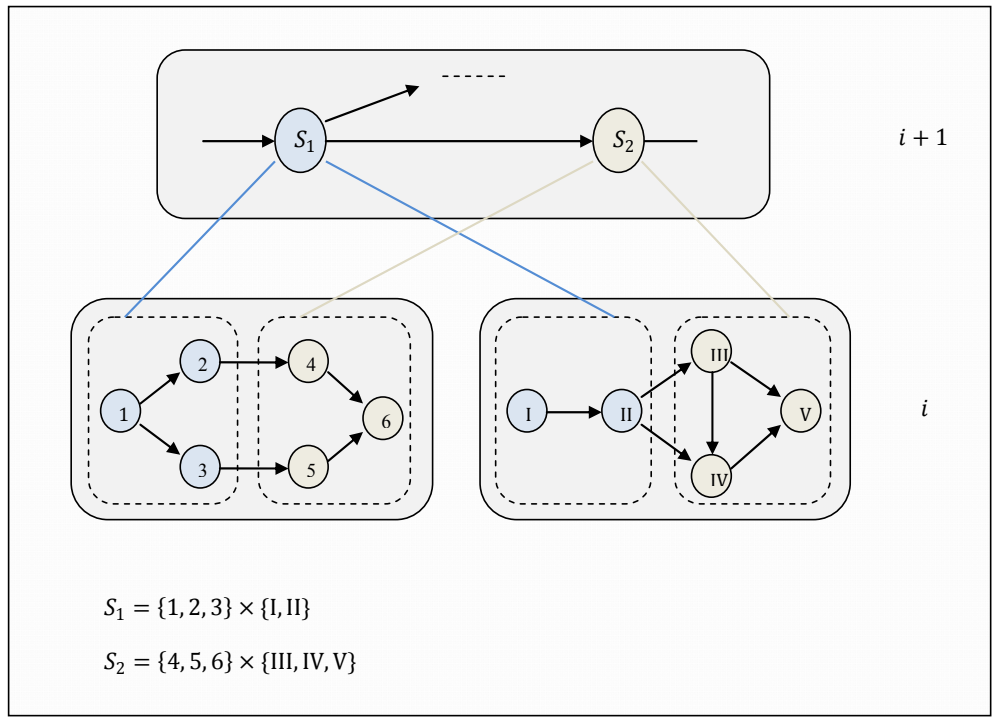

**Figure 15:** Aggregation of state diagrams - children dynamical systems ($i$), parent dynamical system ($i + 1$).

The composition of state diagrams allows one to formally represent different combinations of complex criteria sets in order to perform objects classification and to solve control problems. Using the consistency rules and operations with state diagrams, one can model diverse schemes of inter-level relations and influence of states of lower level diagrams on the processes at higher levels of hierarchy. As a result, certain value is produced at the output of the highest level; this value is considered as a response of the hierarchical system on the input control symbols.

**5.4. Model of Controllable Development of System**

Elementary and complex state diagrams enable one to construct clear and graphical development models. The nodes of diagrams are states, and the arcs are intensities of objects transitions from one state to another. The ordering of states demonstrates the process of objects development. Using the modeling tools, the development model can be redefined and new states and new ordering relations can be added. Considering the state diagrams in time domain (time-domain analysis) - animation of objects' flows - allows forming the time characteristics of the process of development of a set of objects under study.

The various approaches to the control processes require consideration of development models, in which the representation of controllable dynamics of hierarchical object initiated by input signals comes to the forefront. The model of controllable development is based on the following principles:

- Selecting the control actions that influence the controllable system; this is important for autonomous construction of control scenario and for flexible modification of the model to the alternative control scenarios;

- Taking into account the states that have been attained on the previous control stages (system state history); this provides a succession of multistage control scenarios;

- Comparing with the results of alternative control scenarios; this provides basic arguments upon assessment of the efficiency of control scenarios.

The model of controllable development illustrates the key dynamic characteristics depending on whether or not the control actions corresponding to the current states are performed. In this sense, the model of controllable development is constructed in the form of "what if" hypothesis.

A hypothesis of controllable development is defined as the tuple

$$H = (S, P, S_0, S^*, X, T)$$

where

$S$: the set of states;

$S_0, S^*$: the initial and final states respectively;

$X$: the set of input control symbols;

$P$: the set of arcs, $P = P^+ \cup P^-$, $P^+ \cap P^- = \emptyset$; $P^+$ is a subset of arcs that define state transitions initiated by input control symbols, $P^-$ is a subset of arcs that define state backsteps when neither of control symbols enter the input;

$X \leftrightarrow P^+$: the one-to-one correspondence that defines for each input control symbol the state transition initiated by this symbol;

To model and analyze the interconnection between state diagrams at different levels of hierarchy we introduce to the model the rules of inter-level dependencies. These rules can be considered as a kind of win/loss scheme that formally describes what one agents (objects) can obtain (benefit, profit), depending on the states of other agents. So, we have

$T$: the inter-level connection rule between state transitions at neighbor levels of hierarchy. The rule is determined by the mechanism of after-effect by means of splitting $P^+$ and $X$ into two subsets $(P_z^+, P_u^+)$ and $(X_z, X_u)$, respectively. The arcs of $P_z^+$ are called isolated, and the arcs of $P_u^+$ are called coupled. According to this partition, control symbols of $X_z$ are called individual, and control symbols of $X_u$ are called general. The coupled arcs are defined by introducing the parent-arcs as a Cartesian product of child-arcs for state transition diagrams at neighbor levels of hierarchy. The isolated arcs $P_z^+$ describe the state transitions initiated by individual input symbols $X_z$; this kind of symbols do not influence the state transitions of other diagrams. The coupled arcs $P_u^+$ describe the state transitions initiated by general input symbols $X_u$; this kind of symbols initiate the state transition on the parent-arc, which means, as a consequence, the initiation of state transitions on the corresponding child-arcs. And conversely, state transitions on all/several child-arcs initiate a state transition on the parent-arc of the diagram of the higher level of hierarchy.

The model of control scenario that manages the development of objects in complex hierarchical system is represented as

$$C = (\Omega, I, F, \vartheta, T)$$

where

$\Omega$: the set of state transition diagrams that describe the state dynamics - development plans - for each object;

$I$: the hierarchical structure identifier; in general, the identifier is related with the topology of the system; this is done by extending the model of control scenario by some variable $\theta$ that will define the system topology and, as a consequence, the structural relations in the system, for example, hierarchical structure, networked structure, *etc.*;

$F : I \rightarrow \Omega$: the functional that assigns the hierarchical number to each diagram from $\Omega$;

$\vartheta$: the time diagram for control symbols $X$; it determines the sequential-parallel process of input control symbols entering;

$T$: the scheme of after-effect given by the inter-level connection rule between state transitions.

The time diagram $\vartheta$ can be given by the use of different ways, including the estimation rules of each current state of the system.

The trajectory of attainable states represents general and local goals solved by control scenario on arbitrary time interval. Several properties can be considered to evaluate the quality and efficiency of scenario:

- Scenario is called *complete* if all subsystems move to the final states of the corresponding state diagrams;

- Scenario is *redundant* if both individual and general symbols, which are coupled in the "parent-child" hierarchy of state diagrams, enter the input of subsystem;

- Scenario with *omitted possibilities* exhibits high frequency of transitions on the arcs that stand for the backsteps of the states already reached;

- *Complexness* of scenario is estimated by frequency of transitions on coupled arcs.

Generally, the study of basic properties of control scenario is based on the analysis of the trajectory of attainable states and its comparison with the expected or predicted effect.

### 5.5. Dynamic Classification of Control Objects

In section 5.2. we pointed out that upon constructing the canonical model the rules of classification are used. Indeed, the use of hierarchical state diagrams technique supposes the initial classification of control objects over the system state space in order to build the canonical model of state dynamics of a set of objects under study. The rules of classification are represented in the form of matrices with production rules as elements, which are the formulas of some language $L$. An element $(i, j)$ of classification matrix, where $i$ is a parameter and $j$ is a class of objects, contains a formula that determines the current process of dynamics of $i$-th parameter changes at time interval. The matrices of this type give rules of one-level classification. However, the rules of multilevel classification are more important in hierarchical models. Multilevel classification is based on the gradual specification of conditions that should be satisfied by objects from a class. In general case, the rules of multilevel classification are heuristic, and reflect the knowledge and experience of problem domain experts. The subclass of multilevel classification rules which along with the grouping of objects reflect the semantics of state dynamics is of the most interest. To define multilevel classification, the notions of state scale and classifier are introduced [120].

Let $K = \{k_1, k_2, \ldots, k_q\}$, $q \leq card$(set of parameters), be a set of predicates related to the values of parameters of a set of objects. Then the ordered set of propositions $K = \{K_1 \prec K_2 \prec \cdots \prec K_n\}$, $n \leq card(S)$, $T_{K_i} \cap T_{K_j} = \emptyset$, $i \neq j$, where $T_{K_l}$ is the truth domain of $K_l$, is called a *one-level scale* if each $K_l$ defines the state $S_l$. It is assumed that propositions and the corresponding states have the same ordering, that is, if $K_1 \prec K_2 \prec \cdots \prec K_n$ then $S_1 \prec S_2 \prec \cdots \prec S_n$. The scale determines the values of parameters and allows one to compare the states of the objects.

The scale $\{K_{i_1} \prec K_{i_2} \prec \cdots \prec K_{i_n}\}$ is said to be the hierarchical continuation of the scale $\{K_1 \prec K_2 \prec \cdots \prec K_i \prec \cdots \prec K_n\}$ if the propositions $\{K_{i_\nu}, \nu = \overline{1, n}\}$ are the set of sub-propositions of $K_i$.

The hierarchical system of scales is referred to as *classifier* for a set of objects over the set of parameters at time interval $\Delta$. The classifier is then used for formal description of state dynamics of the objects from the set. The classification of objects over the states allows us to construct the state space of the system.

**5.6. General Algorithm for Construction of System Development Model and its Analysis**

We outline a general algorithm for construction of development model for objects at one level of hierarchy. The algorithm is a basic algorithmic process for building the development model of complex hierarchical system with polymorphic parameters [120].

Let us denote by $W$ the set of objects of one level of hierarchy. The algorithm is divided into four stages.

**Stage 1:** includes the preliminary study; at this stage one should establish the parameters which dynamics goes in parallel and which characterize an object from $W$. At first stage we choose the set of parameters and draw the graph representation of their parallel dynamics at a given time interval. Using a graph representation, we compare the character of parameter value changes of the object.

**Stage 2:** the stage of estimation of dynamics of the parameters; it consists in getting the comparative dynamical characteristics of polymorphic parameters for different objects from $W$, and in extrapolating the dynamics of parameter values for arbitrary object with simple relationships, which describe the essence of processes that are being studied. The analysis of parameter dynamics gives answers to the following questions: whether a parameter is a function of time of any standard type - monotone increasing or decreasing, with one or several critical points, whether the function is bounded, whether it has a point of inflexion, or it can be described by a cyclic process.

**Stage 3:** the stage of recognition of the type of dynamic processes; it consists in estimation of states of parameter dynamics. This includes heuristic analysis of a sequence of parameter values $X(1), X(2), X(3), \ldots$ at discrete time moments, which then produces the estimate of the current state at arbitrary time moment $t$, $S(t) = F(S(t-1), X(t-1), X(t))$, where $F$ is a function, $S(t-1)$ and $X(t-1)$ are respectively the state and the value of parameters at previous time moment $(t-1)$. This heuristic process is universal and applicable for every parameter, for which values the operation of comparison is defined. A qualitative estimation of the current process of dynamics of the parameters allows one to create diverse classification rules $K$.

**Stage 4:** at this stage the classifier determined by the classification rules $K$ is used for formal description of dynamic development model of objects from $W$. Formally, the description of the dynamical system is given in the form of canonical model of development of objects from $W$.

Canonical models along with the actual state dynamics allow one to get a qualitative picture of the development processes of the dynamical system under study.

These four stages of the algorithm comprise the general scheme of study of a set of objects at arbitrary level of hierarchy as a unified dynamical system.

### 5.7. Criteria for Efficiency of Control Scenarios

One can see that the method of HSGD formalizes the state dynamics processes in the form of mathematical mappings. In a generalized form, the model of system development and hierarchical control can be represented as follows.

Let $U_{ij}$ be a set of control actions for $ij$-th subsystem, $U_{ijk} \subset U_{ij}$ be a subset that corresponds to the $k$-th state, and let $\Delta$ be the control time interval and $u(i,j,t) \in U_{ij}$ the control action on $ij$-th subsystem at time $t \in \Delta$. Then the control process is formally described by the vector-function

$$u(i,j,t) = (u_1(i,j,t), u_2(i,j,t), \ldots, u_n(i,j,t))$$

in control space $\prod_{i=1}^{n} U_i$ of Cartesian product of sets of control actions on different subsystems. It is assumed that $u(i,j,t)$ has a unique influence on subsystem state and on the value of its efficiency criterion.

Then, let $s(i,j,t)$ be the process of state changes, and $w(i,j,t)$ be the process of changes of efficiency criterion for $ij$-th subsystem at control time interval $\Delta$, $t \in \Delta$. Then, the vector-functions

$$s(i,j,t) = (s_1(i,j,t), s_2(i,j,t), \ldots, s_n(i,j,t))$$

and

$$w(i,j,t) = (w_1(i,j,t), w_2(i,j,t), \ldots, w_n(i,j,t))$$

formally describe the attainable configurations that represent the efficiency of control process $u(t)$ at time instant $t \in \Delta$.

The above vector-functions can be used to define the concrete efficiency criteria in accordance to the needs of simulation and priorities of decision-maker.

### 5.8. Modeling System States and Process of Analysis

The tool of state diagrams and HSGD method enable one to describe the process of development of complex system in highly information-intensive environments. This information (problem) environment is broken down into a collection of problem domains which represent the information levels of the system. Each information level consists of a number of information contexts. The concept of information context allows one to combine in one model a large amount of diverse information of different character (Fig. (**16**)). This helps one to explore the model in different aspects and, based on the knowledge obtained by simulation, to synthesize a holistic view on the system. Based on the agent oriented model of the system and object oriented approach to the control and problem domains, each information level describes the system as a discrete-continuous dynamic multilevel model, which reflects the process of state changes of the system in the phase space. Discrete properties of each of the information levels of the system are determined by the necessity to divide the state space of each of the objects into subspaces in order to reflect the observations that characterize the change of states of the objects upon transition from one subspace to another. Our approach uses a combination of analytical and empirical relations with the methods and means of artificial intelligence, where the available analytical and empirical relations are supplemented by formalized or weakly-formalized expert knowledge.

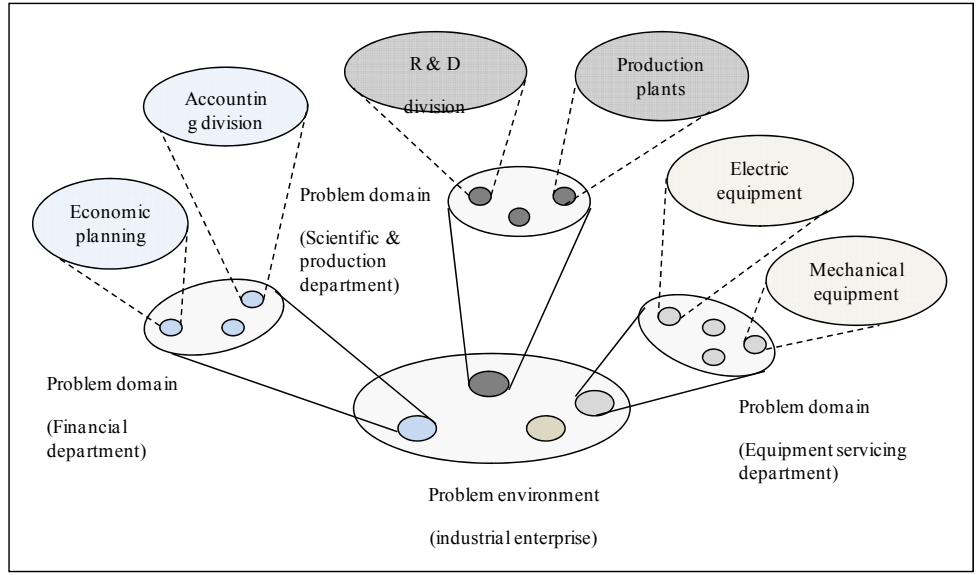

**Figure 16:** The demonstration of dividing of problem environment into information levels and information contexts: problems environment - enterprise, problem domains (information levels) - financial department, scientific & production department, *etc.*, information contexts - economic planning division, accounting division, R&D division, production plants, *etc.*

The state transitions are described by the information-mathematical model provided by the combination of data and knowledge stored in databases, knowledge bases, in the existing ontologies and mathematical modules. The subspaces of states are defined by means of the classifier. Using the selected information levels and contexts and the subspaces of states, the overall process of modeling system states and its analysis is formally presented as follows:

$$M = (\Delta, S, X, Mn, U, Y, V, V_y, Q)$$

where

$\Delta \subset \mathbf{R}$: the finite time interval of system modeling;

$S$: the state space; $s = (s_1, s_2, \ldots, s_n)$, where $s_i$ is the phase coordinate. Each $s_i$ can be bounded from both sides, $\underline{s_i} \leq s_i \leq \overline{s_i}$, $\underline{s_i}$ and $\overline{s_i}$ are the restrictions on the values of phase coordinates, $i = \overline{1,n}$; $s(t)$ is the phase trajectory of state changes. The phase coordinates describe the properties and states of the system on different information levels. The set of phase coordinates has an object-oriented structure, that is, $s_i = (s_{i1}, s_{i2}, \ldots, s_{ir})$, each coordinate is described by a collection of its own properties - factors. The phase space $S$ is represented as a partition $\prod_{j=1}^{k} S^j$ that corresponds to the hierarchical structure of state changes and incorporates the notion of hierarchical state, $k$ is the number of nesting hierarchy levels.

$X$: the set of admissible inputs, $x = [x^1, x^2, \ldots, x^l]$, $l$ is the number of inputs, each input accepts information of its type that corresponds to the information contexts;

$Mn$: the set of output results $m$ of monitoring and/or diagnosing of control object, $m = [m^1, m^2, \ldots, m^p]$.

$U$: the set of control actions, $u = [u^1, u^2, \ldots, u^q]$. The control action is determined by the causality "current state - control action", needed to steer the objects to the desired state with prescribed values of parameters; control actions can be formalized by means of production rules.

$Y$: the set of output information that characterizes the values of parameters that define the states of objects, $y = [y^1, y^2, \ldots, y^h]$.

$V = (V_x, V_m, V_u, V_h, Im, Pr)$: the transition operator that defines the set of current states, based on the results of the previous modeling step (prehistory), with some priority value $p \in Pr$, that is, $(s^w(t), p^w) = V(s(t-1), t, p)$, where $w = \overline{1, W}$, $W$ is the number of current states under study; $V_x$ is the operator of synthesis of new initial conditions and new states that depends on the input information, $V_m$ is the operator of generation of information on the results of monitoring and/or diagnosing, $V_u$ is the operator of formation of the control actions, $V_h$ is the operator of formation of information from lower level of hierarchy, it uses the inter-level connection rules $T$ - the mechanism of after-effect; $Im = \{Db, Kb, Ma, O, PR, p\}$ is the information-mathematical model that describes the behavior and state dynamics of objects between the events, $Db$ - the databases, $Kb$ - the knowledge-bases, $Ma$ - the analytical models of state dynamics of objects at time intervals, $O$ - the ontologies of problem domains (information levels), $PR$ - production rules constructed by experts and based on $Db$ and $Kb$, which give rules for re-estimation of current states around the dynamics of parameters and serve for possible generation of new states; $p$ is the priority of the behavior of state dynamics of the objects, which characterized by expert knowledge and decision-maker;

$V_y$: the output operator, $y(t) = V_y(s(t-1), t)$, that transforms the information on the state of the object to the values of parameters to be controlled on the next step;

$Q$: the qualitative characteristic of the system state; it is determined by the combination of expert estimates of this state and the profit of reaching this state - what goals are achieved; it depends on the priorities $p$ of the previous states.

The possibilities provided by the information-mathematical model and its analysis are the tool for decision support. The simulation process on the above models allows one to predict the changes in values of parameters over all problem domains of the system and predict future system states, and thus justify the decisions to be made.

Finally, we would like to outline some directions for further development of HSGD tool:

- Developing probabilistic model of HSGD and methods of its analysis;

- Describing HSGD components and dynamics by means of mathematical functions (operators) to get rigorous mathematical analysis of the model;

- Studying HSGD as hierarchical dynamical network, using mathematical methods and means of dynamical systems theory;

- Describing HSGD in hierarchical game-theoretic setting and developing the methods of its analysis;

- Modeling and control of conflict situations with conflict goals at different hierarchy levels;

- Modeling and control of system development under resource- and time-limited settings (or, some other interested criteria), using methods of optimal control theory;

- Building heterogeneous and distributed space-time HSGD models;

- Structural graph-theoretic analysis of HSGD: identification of unwanted cycles, stability analysis, robustness of given situations, *etc.*;

- Applying HSGD for modeling and solving software engineering problems, and in some other areas, for example, social sciences, ecology, *etc.*

Some improvements, extensions, or new variables and functions may be included in the current model of HSGD to get an adequate model for the above directions.

## 6. INTELLIGENT INFORMATION SYSTEM FOR SIMULATION AND DECISION SUPPORT

Computer information system intended for large scale simulations and decision support should inevitably be based on the comprehensive use of information technologies [147, 148]. The system should have a modular structure that provides sufficient convenience and facility of editing of the separate modules, not influencing the functioning of the others, and adding of new functional capabilities. If it is possible for application domain, the information system should be built in the form of master-system that could be able to dynamic adjustment to the specific problem domain, and in order the information environment of the system could be adapted to the current range of problems.

### 6.1. Characteristics of Intelligent Simulation Environment

Computer simulation models that provide intelligent simulation environment within the information system should meet the following main requirements:

- *Model completeness*. The models should provide sufficient possibilities for obtaining the necessary characteristics of system with the required accuracy, reliability, and confidence;

- *Model flexibility*. The models should enable one to reproduce various situations upon changing of system parameters;

- *Model structure*. The models should provide the possibility of modification of their separate parts;

- *Information support*. It should provide the information compatibility of models with computer databases.

An intelligent simulation environment is a large knowledge-based integration system, which consist of several symbolic reasoning systems (LISP, PROLOG, *etc.*) and numerical simulation software. Such environments suggest a framework for integration of numerical simulation, expert system and artificial intelligence techniques. In a goal-oriented environment, once the system is described and the goals specified, the simulation system drives itself to goal achievement. We propose intelligent simulation environment of various control strategies and scenario modeling. Several studies show the applications of intelligent simulation in the area of production systems. Recently, an integrated knowledge-based model is developed for complex man-machine systems [149]. Intelligent simulation environments are also proposed for flexible manufacturing systems, information systems, process plants, *etc.* [150-153]. In our approach, an intelligent modeling environment is a flexible, integrated, and knowledge-based framework capable of extending, learning and correcting itself. It is goal oriented and searches for the best solutions by referring to desired target.

In parametric modeling and analysis, the knowledge of values of various parameters and their dynamics is one of the most important elements that provides adequate representation and modeling of state dynamics of complex systems. Monitoring schemes allow one to carry out the observation for the current values of parameters and for the actual information on the character of system parameter dynamics. This information is then used to evaluate conditions/situations around system and to predict possible events in the system and consequences following from them, which can be caused by changes in values of parameters.

Intelligent simulation environment is proposed by integration of: 1) an integrated database and modeling, 2) rule-based (goal-oriented) behavior and 3) parametric and flexible structures. It is a goal oriented, flexible and integrated approach and produces the optimal solution by referring to integrated models and knowledge- and databases. The properties and modules of the prescribed intelligent simulation environment are: 1) parametric modeling, 2) flexibility, 3) integrated modeling, 4) rule-based module, 5) integrated knowledge- and databases and 6) learning module.

Generally, the information on the problem domain and properties of system is contained in specialized databases; knowledge about parameters and processes is contained in knowledge bases; information about the current values of parameters and on the character of state dynamics of object is contained in monitoring databases. The investigations on the models are provided by the combination of the methods of (1) dynamic knowledge-based expert systems, (2) production expert systems, (3) database processing techniques, (4) monitoring/ diagnostic data analysis, (5) scenario control and modeling.

### 6.2. Characteristics of Decision Support System

A Decision Support System (DSS) is an information system that supports decision-making and system management activities [154-157]. A properly-designed DSS is an interactive software-based system intended to help decision makers in compiling useful information from raw data, documents, personal knowledge, business models, *etc.*, to identify and solve problems and make decisions. A DSS enables the user to make fast, responsive decisions using all the necessary information appropriate to the task at hand.

According to Sprague and Carlson [157], a basic DSS framework contains a database and model base, with associated management systems, and a dialog component through which the decision maker interacts. Other features include graphical-based dialog interface (a graphical user interface, GUI), output facilities and report generators, and event logging. Haag *et al.* [148] describe these three components in more detail. The Data Management Component stores information (which can be further subdivided into that derived from an organization's traditional data repositories, from external sources, or from the personal insights and experiences of individual users); the Model Management Component handles representations of events, facts, or situations (using various kinds of models, two examples being optimization models and goal-seeking models); and the User Interface Management Component is, of course, the component that allows a user to interact with the system. Levin [158] analyzes a number of works and names the following components as essential for a modern DSS: (1) models, which include multi-criteria techniques, problem-solving schemes, data processing and knowledge management; (2) analytical and numerical methods of data pre-processing and identification of problems for the preliminary stages of decision making; (3) human-computer interaction and its organization through graphic interface and others; (4) information support, communication with databases, web-services, *etc.* According to Power [155], academics and practitioners have discussed building DSS in terms of four major components: (a) the user interface, (b) the database, (c) the modeling and analytical tools, and (d) the DSS architecture and network. The definition of a DSS, based on Levin and Power, in that a DSS is a system to support and improve decision making, represents, in our opinion, an optimal and suitable to our needs background for information DSS requirements. The traditional decision making workflow consists of several standard steps (Fig. (**17**)).

The DSS structure has to satisfy the requirements, imposed by specialists, and characteristics and restrictions of the application domain. The decision making process includes the preparatory period, situations simulation, analysis of influence of control scenarios, the development of decisions and, finally, the decision making itself and its realization.

As a rule, characteristics of successful DSS experiences include: assistance in semi-structured decision tasks, support of managerial judgment, useful to non-computer specialists working interactively with the system, exploits both models and databases to generate information, and adapts to the decision-making approach of the user. A DSS does not necessarily look for optimal solutions to complex and/or operational problems, but rather as Ackoff [159] states, a DSS supports "a decision for which adequate models can be constructed but from which optimal solutions cannot be extracted". The levels of automation and support to the decision maker can vary from high to low. Strong [160] found that high levels of decision automation and high levels of decision support provided a best DSS environment to support decision makers. Other approaches and techniques for successful decision making and analysis have also been proposed [161-165] recently.

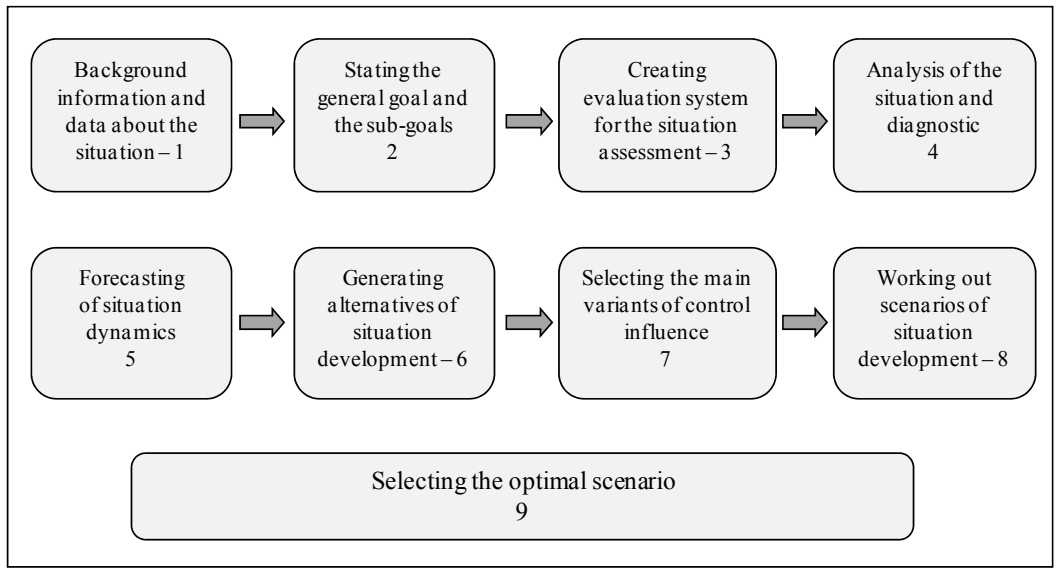

**Figure 17:** The traditional scheme of decision-making.

### 6.3. Architecture of Intelligent Information System

The architecture of information system consists of several independent but logically related to each other subsystems [128, 129]. The subsystem of *Direct Planning* is intended to establish complex problem points, to cope with a collection of interconnected complex control problems, to construct control scenarios for solving complex problems and to compare them in accordance to the efficiency criteria. The subsystem includes the following components (Fig. (**18**)):

- *User interface* provides interaction of user with the system. The software of user interface serves as a tool that realizes all functions of computer system associated with information exchange with the user.

- *Library of parameters* contains blocks of parameters for supporting the continuous process of observing for a number of parameters; the library is extendable and editable.

- *Dynamic Knowledge-based Expert System* is a computer realization of formalized expert knowledge about problem and control domains, and it is used in building dynamic models.

- *Builder of canonical model of state dynamics* is a specialized module of entering of state diagrams as input information. The state diagrams tool provides clear and precise formalization of states, inherent for one process and not typical for others. It can be used for representation of regularities and typical models of state dynamics.

- *Monitoring databases*. Direct planning in control systems assumes a high level of informatization and operative connection with monitoring database.

- *Interpreter of monitoring database*. The interpreter of monitoring database and the model of controllable state dynamics are the basic components in automated computer information system. The interpreter of monitoring database operates according to the composition of canonical models of state dynamics specified by the user and generates the description of actual multilevel dynamics of the object.

- *Model of controllable state dynamics* is constructed as expert subsystem for qualitative estimation of control scenarios defined by the user.

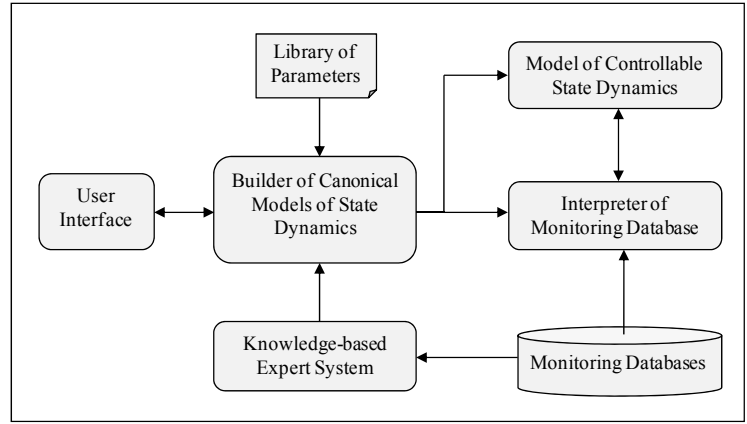

**Figure 18:** A structure of direct planning subsystem.

Some explanation should be given for the component *Dynamic Knowledge-based Expert System*. We wish to note that one of the approaches to the synthesis of dynamic models of complex systems is based on the use of the so-called *master-systems* consisting of canonical templates and expert knowledge-bases. The knowledge-base consisting of declarative and procedural knowledge realizes a conceptual model of complex system. The declarative knowledge contains:

- Objectives tree of complex system that provides a decomposition of global goal on subgoals and description of relation between them;

- The architecture and/or structure of complex system;

- The set of canonical templates;

- The set of models of canonical templates;

- Problem domain databases.

The information about objectives tree can be represented as the tuple

$$O = (I, Id, G, L)$$

where $I$ is a structure that determines the decomposition of global goal, $Id$ is the structural identifier for nodes, $G$ is a goal assigned to the node, $L$ is a rule/law that describes the connection between the neighbor nodes.

The canonical template can be realized with the use of the language of system dynamics [47] or state transition diagrams. The canonical template has a certain structure, a set of input, output, and initial values/conditions. Formally the canonical template can be described as

$$C = (Str, T, X, Y, U, Iv, Tr)$$

where $Str$ is the structure of the template, $T$ is a rule/law of template functioning/behavior, $X$ is a set of input parameters, $Y$ is a set of output parameters, $U$ is a set of control symbols, $Iv$ is a set of initial values/conditions, $Tr$ is a set of rules that govern the transformation of template structure, which means adding, modifying, removing links and/or nodes.

The canonical template is a separate object having basically information about its components and certain internal structure. But the canonical template model is the object that may contain information not only about its components and structure but also about the concrete input and output values and values of initial conditions. Each canonical model is assigned to one of the goals of the objectives tree.

The synthesis of dynamic models of complex system is realized by transformation of declarative knowledge about problem domain to the algorithms of system state dynamics with the help of procedural knowledge.

The procedural knowledge is realized in knowledge-bases in the form of inference rules. They formalize the process of dynamic models synthesis. The inference rules provide the mapping of structure of conceptual model into the structure of dynamic models. The knowledge-base can contain different groups of inference procedures, depending on the purposes of the investigation. For example, they can be of the following types: correspondence rules that determine for each canonical model the goal problems it solves; the inference rules that define informational relations between the templates in canonical model, *etc.*

The representation of conceptual model of complex system in the form of knowledge-bases provides the autonomous usage of expert knowledge upon synthesizing the dynamic models. The above model is extended by adding to the canonical templates a set of input control symbols, thus providing the dynamic models with the mechanism of system control.

So, the *Dynamic Knowledge-based Expert System* is constructed in the form of master-system, it dynamically adjusts to the specific problem domain requested by the user through the Interface, provides corresponding templates for synthesis of models, connects with the appropriate database, and provides the relevant to the selected problems knowledge-base.

The subsystem of *Direct Planning* realizes the two stage process of simulation and control (Fig. (**19**)). The first stage is the stage of retrospective analysis consisting in the construction of predictive model of inertial state dynamics, which is the model of dynamics of object when no control actions are undertaken. The retrospective analysis provides the means for event prediction. When predicting events, the parameters of system are continuously measured. If there was some event in a system and for some time before the event a parameter has sharply changed, or there was a gradual change of values of parameter up to some critical, then such anomaly is related with this event. The dependencies of such kind confirmed repeatedly, *i.e.* becoming stable, are used for estimation and prediction of possible future events in the system. Actually, knowledge and experiences obtained in the past and expert knowledge are used. The stage of retrospective analysis provides user with the tools of selection of objects, study of chosen parameters, and construction of state diagram, which interprets the monitoring data. This implements the diagnostic analysis of objects' state dynamics. By analysis, experimentation, and selection of different sets of parameters, the diagrams that most expressively depict "negative" (positive) trends are found. These diagrams formally represent the current control problems and answer the question "what will be if no control actions are performed". The results of retrospective analysis help put forward the goals and control problems and help one to form possible alternatives of controllable state dynamics for the perspective period.

The second stage consists in construction of the model of controllable state dynamics which includes statement of control problem, alternative control scenarios, expert estimation and selection of control scenarios of object's state dynamics. The stage includes the construction of the model of controllable state dynamics, analysis of controllable processes of objects' state dynamics, and obtaining the answer to the question "what control actions should be undertaken to achieve the required goals". At this stage, an initial state and expected final state are described and the space of intermediate states is constructed. Then, the conceptual model of control scenario in the form of States Generator is defined.

The subsystem of *Multilevel Dynamics Simulation* includes the components (Fig. (**20**)):

- *System of canonical models of state dynamics* of hierarchical object provides the necessary tools for construction of canonical state dynamics models of simulated objects.

- *Scheme of coordination of canonical models* is based on the rules of composition of state diagrams.

- *Description of actual state dynamics* of hierarchical object uses the monitoring databases that store the information on the actual dynamics of parameters.

- *Monitoring databases* contain the current information about the dynamics of parameters, which is the result of monitoring or diagnosing of the object.

- *Partial/incomplete canonical m*odel of state dynamics. The model of controllable state dynamics is used for checking a hypothesis about the efficiency of scenario being estimated. The scenario estimation criterion is given in the form of "partial" or "incomplete" state dynamics diagram determining support states that should be reached with the specified restrictions on time and resources.

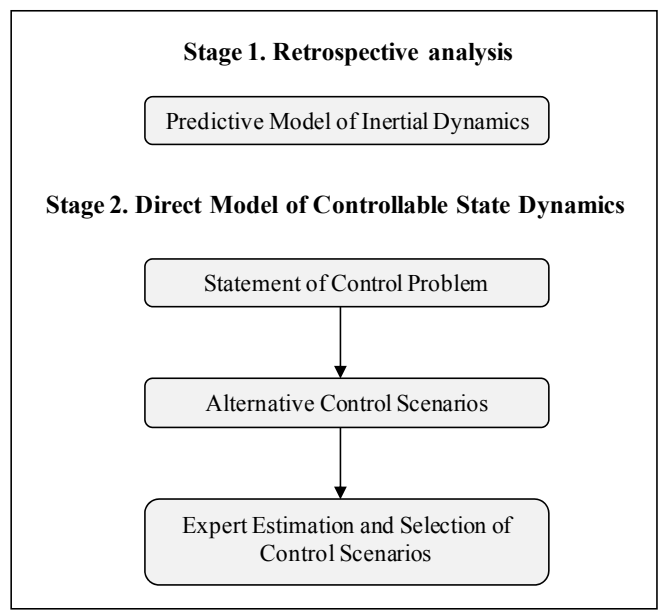

**Figure 19:** Generalized scheme of the process of direct planning.

The special case of "partial/incomplete" state dynamics diagram is the pair of states: initial state and desirable final state. In this case, the expert subsystem should: either confirm a hypothesis that the model of state dynamics, controlled on the basis of the scenario, meets the given criteria or requirements, and supplement the input diagram with the specifying intermediate states, or refute the hypothesis and generate computer prediction in the form of alternative state dynamics diagram.

- *Scenarios of state dynamics/Conceptual model of scenarios* - States Generator. The Scenarios of State Dynamics serves as an inference system. It is considered as generator of consecutive states of object under investigation. The rules of States Generator are represented in the format of tree-like decomposition of global goal on the sub-goals; to each terminal node an elementary rule is assigned. The rules are represented in the IF-THEN format (Fig. (**21**)):

$$(S_i, S_k, S_l, u_{ik}, r_{ik}, t_{ik})$$

where $u_{ik}$ is the control action needed to be undertaken for an object to be steered from the state $S_i$ to the state $S_k$ with the use of resources $r_{ik}$ and in time $t_{ik}$, while not admitting the backstep to the previous state $S_l$.

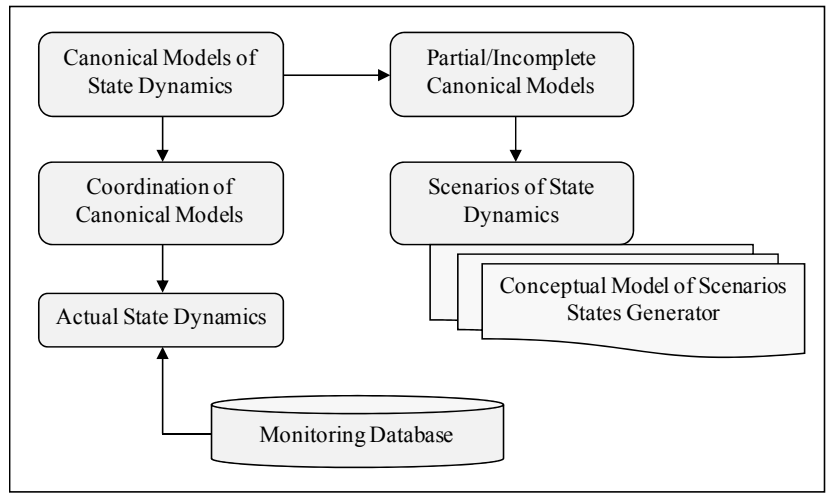

**Figure 20:** Structure of Multilevel Dynamics Simulation Subsystem.

The construction of scenarios is divided into several steps: (1) analysis of initial state of object and of possible trends of state's changes, (2) exploration of a spectrum of possible future states of the object, (3) formulation of hypotheses about tendencies in transitions from those states to the subsequent ones, (4) analysis and formulation of desired end results - final state of the object.

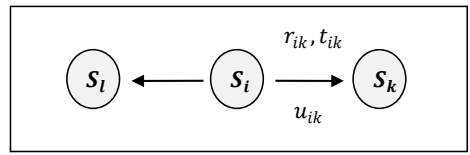

**Figure 21:** The format of elementary IF-THEN rule.

The subsystem of *Simulation Process Control* includes two modules (Fig. (**22**)), *Executive Module* and *Control Module*:

- *Executive Module* consists of database of objects' control models, database of state dynamics simulation models, database of analyzable control actions, simulation system of controlling scenarios, and database of complex control actions. Simulation system of controlling scenarios that uses the database of objects' control models, the database of state dynamics simulation models, and the database of analyzable control actions provides a tool for formal representation of goals, control problems and state dynamics, time and resource characteristics.

  The simulation system of controlling scenarios enables user to analyze global efficiency of control actions directed on the achievement of goals at different levels of hierarchy. The system allows user to not just compare separate control actions but also to formally synthesize complex control actions (control strategies) for hierarchical system as a whole, and then provides their further comparison. The functional sub-modules of Executive Module reproduce the stages of control actions selection, including the stage of automatic synthesis and initiation of problem domain and control models.

- *Control Module* realizes the functions of expert estimation and comparison of control action sets, knowledge-base support, and decision making.

The simulation process control subsystem provides the tools of synthesis, simulation, and analysis of control scenarios.

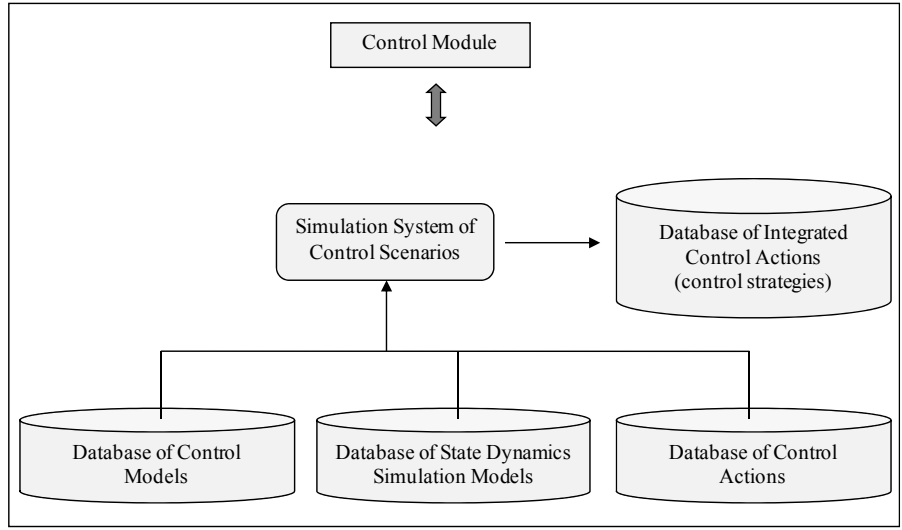

**Figure 22:** Structure of Simulation Process Control Subsystem.

The rules of state transformations and the scheme of construction of control scenarios: (1) allow one to easily realize iterative process of creation and modification of control scenarios, (2) admit the efficient realization by means of executive procedure, (3) possess the sufficient expressiveness of specification of control processes.

The State Generator enables user: (1) to study effects of integrated and multi-aspect control regarding different objects of complex system, (2) to divide the control process into stages, (3) to perform decompositional schemes of prediction, in which each subsequent model is an integrated or detailed elaboration of the previous, (4) to construct and analyze the interconnected aggregated and detailed models of state dynamics of the system.

The flexibility and comprehensiveness of the architecture of information system provide the implementation of the following functions:

- Identification and registration of the information about the events and the current situation around the complex system. Information and expert knowledge storage, maintaining and management;

- Description of system parameters and those of monitoring and/or diagnosing;

- Description of control actions;

- Description of information-mathematical model;

- Description of the structure of state space;

- Description of the structure of control and problem domains and informational levels of system dynamic models.

The overall architecture of computer information system for simulation and decision support is presented on Fig. (**23**).

The general scheme of the process of system simulation analysis and decision making based on the model of HSGD with use of the methodology of hierarchical system scenario control and control strategies consists of five interconnected levels (Fig. (**24**)).

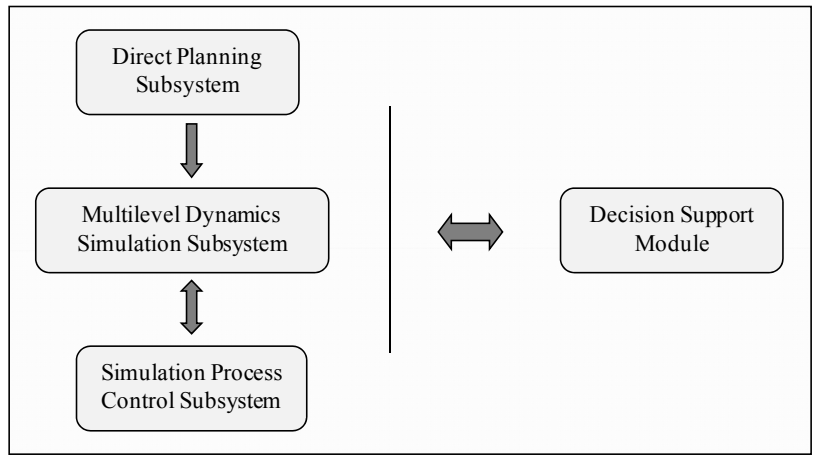

**Figure 23:** General architecture of information system.

The computer information system allows user to:

- Define global goal of complex system and to represent a decomposition on the subgoals, and to describe relations between them;

- Synthesize dynamic models of complex system;

- Determine efficient control actions in solving control problems and achieving system goals;

- Realize the iterative process of creation and modification of control scenarios;

- Study multi-aspect control of different subsystems and analyze integrative behavior of complex dynamic system as a whole;

- Construct and analyze different control scenarios, study and compare them for efficiency;

- Study behavior of system under different initial conditions and to study state dynamics of system for different groups of parameters;

- Get a holistic view on a complex system and its behavior;

- Make integrated and well-founded decisions.

The proposed architecture provides the most complete tools of simulation and decision support in relation to the problems and goals that were outlined in the beginning of this section. This information system is quite suitable for applications in weakly-formalized and ill-structured problem domains. It enables one to:

- Involve modelers and decision-makers in the process of formalization of criteria for scenarios evaluating;

- Provide users with the ability to put the current tasks of problem analysis in a language close to the professional;

- Provide the possibility of considering the problem with various degrees of specification, using operational data of the level requested;

- Provide users with the possibility to conduct long-term computer archives of statistical and monitoring data of specific applied models of systems;

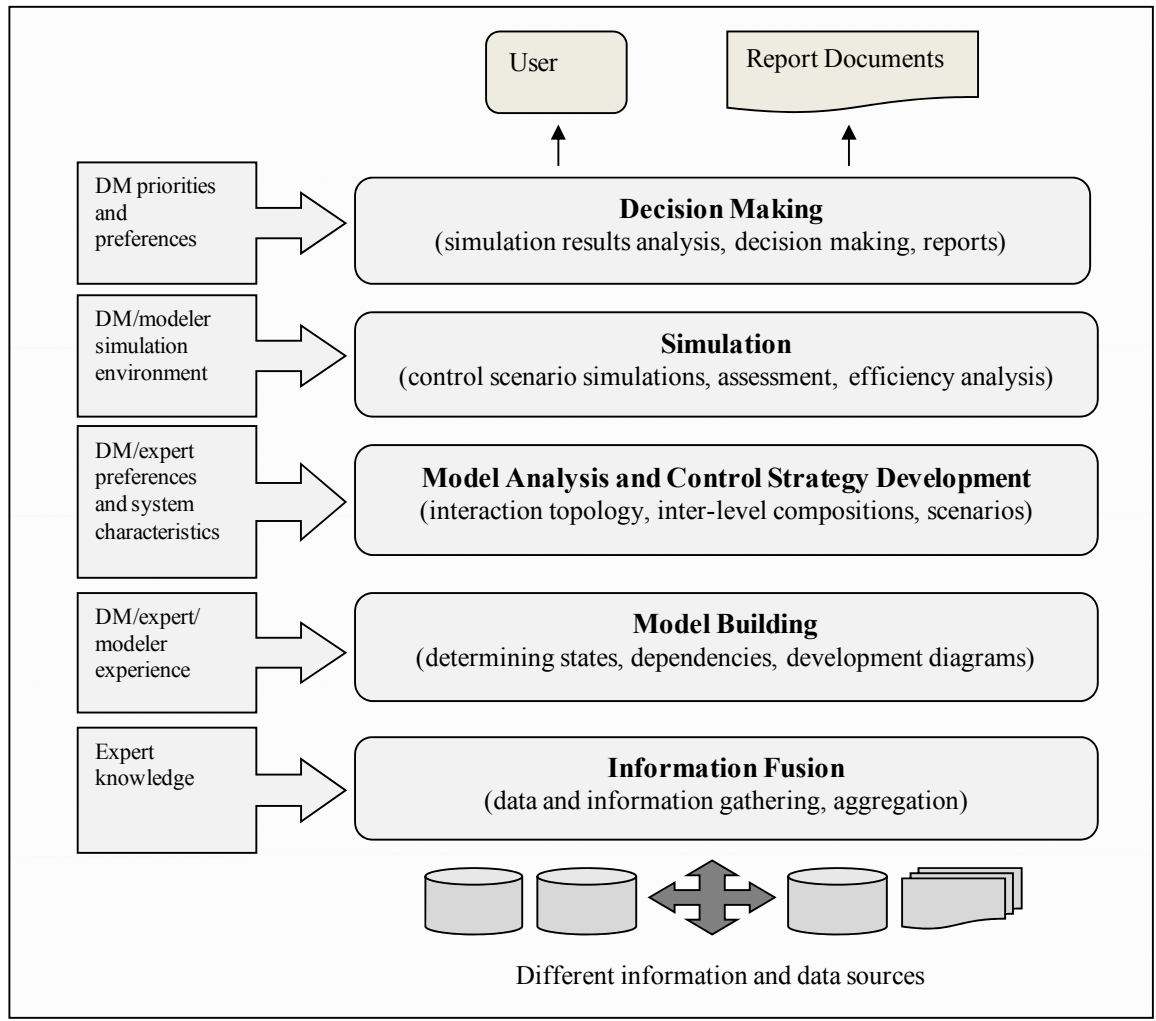

**Figure 24:** The general scheme of system simulation analysis and decision making.

- Accumulate knowledge about the problem and control domain, build and use the information system, open to correction and supplementation;

- Use expert knowledge about the quantitative and qualitative assessments of interrelationships between dynamic system parameters.

The system enables to conduct the analysis of significance of control scenarios in relation to the complex problem solving and provides the ability to integrate it with the actual process of planning and control.

## 7. CONCLUSIONS

To adequately model and analyze processes in complex real-world systems, we need the holistic and integrative methods that combine in them both top-down and bottom-up approaches. It is often not necessary to cope with the all available system parameters and functions in order to get a good approximation of modeled system. Even simpler models lead to very complex behaviors [166] and allow us to obtain sufficiently detailed and comprehensive understanding of the processes in the system. Although it is true for physical and biological systems, for many other kinds of systems, like social, cognitive, engineering, business-process, some systems of artificial intelligence, and others, we have to take into account as many system interrelations, processes, dependencies as it possible. This can only be done by first using methods of systems analysis, and then one needs techniques that have been developed in the field of systems theory and control sciences.

In this chapter we have discussed the problems of modeling, control and decision support in complex systems from the viewpoint of systems theoretic approaches and methods. We considered the basic characteristics of complex systems and analyzed system theoretic approach to complex systems study. We have also considered the main frameworks and paradigms of modern system modeling and control, and analyzed the most widely used methods for mathematical modeling and simulation of complex dynamic systems.

We have proposed the general dynamic modeling technique for complex hierarchical systems consisting of many objects functioning in control loop. The generalized algorithmic schemes for analysis of the proposed models are also presented. The proposed technique uses the object-oriented multi-agent paradigm, and is based on the information-mathematical models and described in terms of the hierarchical state transition diagrams. The method allows both qualitative and quantitative analysis of system behavior and state dynamics through hierarchical scenario calculus. For evaluation of different scenarios the efficiency criteria related to the control strategies are presented. We have also proposed a general structure of computer information system for simulation and analysis of dynamic processes, control strategies and development scenarios in complex systems, which can be used in decision making process.

The technique presented can also be used as a technology for design and building of information systems for simulation analysis of development strategies and control scenarios of complex objects, and it has been applied in several information systems and decision support systems.

The models proposed are especially powerful in information-intensive environments. The information can be simultaneously aggregated in several directions: by hierarchical structure of processes and states embedding, by parallel representation of dynamical characteristics of several processes within one state, and by dividing the observation time interval in accordance to the events associated with the state changes in dynamics of system. The proposed models and technique are sufficiently universal and at the same time they are problem-oriented. The technique can be equally used for various kinds of systems such as technical, engineering, organizational, socio-economic, strategic planning, long-term forecasting systems, and decision support systems.

## REFERENCES

[1]    Y. Bar-Yam, Making things work: Solving complex problems in a complex world. Cambridge, MA: Knowledge Press, 2004.
[2]    Y. Bar-Yam, Dynamics of complex systems. Reading, MA: Addison-Wesley, 1997.
[3]    M. Mesarovic, D. Macko, and Y. Takahara, Theory of hierarchical multilevel systems. New York: Academic Press, 1970.
[4]    M.D. Mesarovic, Theory of hierarchical multilevel systems. In: Mathematics in Science and Engineering. New York: Academic Press, 1970.
[5]    M.L. Estep, Self-organizing natural intelligence: Issues of knowing, meaning, and complexity. Springer-Verlag, 2006.
[6]    H.H. Pattee, Ed., Hierarchy theory: The challenge of complex systems. New York: George Braziller, 1973.
[7]    S. Salthe, Evolving hierarchical systems: Their structure and representation. New York: Columbia University Press, 1985.
[8]    V. Ahl and T.F.H. Allen, Hierarchy theory: a vision, vocabulary, and epistemology. New York: Columbia University Press, 1996.
[9]    R. Badii and A. Politi, Complexity: hierarchical structures and scaling in physics. Cambridge: Cambridge University Press, 1997.
[10]   B. Zeigler, H. Praehofer, and T. Kim, Theory of modeling and simulation. Academic Press, 1999.
[11]   P. A. Fishwick, Handbook of dynamic system modeling. Chapman & Hall/CRC, 2007.

[12]   V. Latora and M. Marchiori, The architecture of complex systems. Santa Fe Institute for Studies of Complexity, Oxford University Press, 2003.
[13]   H. A. Simon, Sciences of the Artificial. Cambridge, MA: MIT Press, 1969.
[14]   N.A. Avouris and L. Gasser, Eds., Distributed artificial intelligence: Theory and praxis. Kluwer Academic Publishers, 1992.
[15]   T.I. Oren and B.P. Zeigler, "Artificial intelligence in modeling and simulation: Direction to explore", Simulation, vol. 50, pp. 131-134, 1988.
[16]   D. Norman and M. Kuras, "Engineering Complex Systems", in D. Braha, A. Minai, and Y. Bar-Yam, Eds., Complex Engineered Systems: Science Meets Technology, Springer, 2006, pp. 206-245.
[17]   B. Blanchard and W. Fabrycky, Systems engineering and analysis. Prentice-Hall, 3rd ed., 1998.
[18]   E. Laszlo, The systems view of the world. New York: George Braziller, 1972.
[19]   M. Mesarovic and Y. Takahara, General systems theory: Mathematical foundations. New York: Academic Press, 1975.
[20]   L. von Bertalanffy, General system theory. New York: George Braziller, 1968.
[21]   S.A. Levin, "Complex adaptive systems: Exploring the known, the unknown and the unknowable", Bulletin of the American Mathematical Society, vol. 40, pp. 3-19, 2003.
[22]   M. Gell-Mann, "Complex adaptive systems", in G. Cowan, D. Pines, D. Meltzer, Eds., Complexity: Metaphors, Models and Reality. Addison-Wesley, 1994, pp. 17-45.
[23]   R. Rosen, "Complexity and system descriptions", in W.E. Harnett, Ed., Systems: Approaches, Theories, Applications. Dordrecht, Holland: D. Reidel, 1977, pp. 169-175.
[24]   R.J. Nelson, "Structure of Complex Systems", Philosophy of Science Association, vol. 2, pp. 523-542, 1976.
[25]   J.M. Ottino, "Engineering complex systems", Nature, vol. 427, pp. 399-399, 2004.
[26]   D.W. Oliver, T.R. Kelliher, and J.G. Keegan Jr., Engineering complex systems with models and objects. McGraw Hill, 1997.
[27]   Z. Bubnicki, Modern control theory. Springer- Verlag, 2005.
[28]   J.L. Casti, "On system complexity: Identification, measurement and management", in J. Casti and A. Karlquist, Eds., Complexity Language and Life: Mathematical Approaches. Berlin: Springer, 1986, pp. 146-173.
[29]   M. Heimdahl and N. Leveson, "Completeness and consistency in hierarchical state-based requirements", IEEE Trans. Software Engineering, vol. 22, pp. 363-376, 1996.
[30]   G. Pappas, G. Lafferriere, and S. Sastry, "Hierarchically consistent control systems", IEEE Trans. Automatic Control, vol. 45, pp. 1144-1160, 2000.
[31]   P. Hubbard and P.E. Caines, "Dynamical consistency in hierarchical supervisory control", IEEE Transactions on Automatic Control, vol. 47(1), pp. 37-52, 2002.
[32]   H. Garcia, "A Hierarchical platform for implementing hybrid systems in process control", Control Engineering Practice, vol. 5(6), 1997.
[33]   P. Albertos, R. Strietzel, and N. Mart, Control engineering solutions: A practical approach. IEEE Computer Society Press, 1997.
[34]   A. Meystel and J.S. Albus, Intelligent systems - architectures, design and control. John Wiley & Sons, 2002.
[35]   I. Dumitrache, "From model-based strategies to intelligent control systems", WSEAS Transactions on Systems and Control, vol. 3(6), pp. 569-575, 2008.
[36]   J.C. Willems, "Paradigms and puzzles in the theory of dynamic systems", IEEE Trans. Automatic Control, vol. 36, pp. 258-294, 1991.
[37]   M. Law and W. D. Kelton. Simulation modeling and analysis. Tata McGraw-Hill Publishing Company Ltd., 3rd ed., 2003.
[38]   N.P. Buslenko, Complex systems modeling. Moscow: Nauka, 1978. (in russian).
[39]   V.V. Rykov. "Aggregative systems: their modeling and simulation", in International conference on Simulation, Gaming, Training and Business Process Reengineering in Operations, Riga, 1996, pp. 38-39.
[40]   V.V. Rykov, Controllable aggregative systems and their simulations. IMACS Trans. Comp. Sci., Amsterdam: North-Holland, 1985.
[41]   C. G. Cassandras and S. Lafortune, Introduction to discrete event systems. Norwell, MA: Kluwer, 1999.
[42]   J. Banks, J. S. Carson, B. L. Nelson, and D. M. Nicol III, Discrete-event system simulation 3rd ed., New Jersey: Prentice-Hall International Series in Industrial and Systems Engineering, 2001.
[43]   K. Wong and W. Wonham, "Hierarchical control of discrete-event systems", Discrete Event Dynamical Systems, vol. 6, pp. 241-273, 1995.

[44] Y.-L. Chen and F. Lin, "Safety control of discrete event systems using finite state machines with parameters", in Proceedings of the American Control Conference, 2001, pp. 975-981.

[45] A.E.C. da Cunha, J.E.R. Cury, and B.H. Krogh, "An assume guarantee reasoning for hierarchical coordination of discrete event systems", in Workshop on Discrete Event Systems, 2002.

[46] R.de Souza, Zhao Zhen Ying, "Intelligent Control Paradigm for Dynamic Discrete Event System Simulation", Discrete Event Dynamic Systems, vol. 9, pp. 65-73, 1999.

[47] J. Forrester, Industrial Dynamics. Portland: Productivity Press, 1961.

[48] Bin Hu, DeBin Zhang, CaiXue Ma, Yong Jiang, XiongYing Hu, and JinLong Zhang, "Modeling and simulation of corporate lifecycle using system dynamics", Simulation Modeling Practice and Theory, vol. 15, pp. 1259-1267, 2007.

[49] M. Ozbayrak, T.C. Papadopoulou, and M. Akgun, "System dynamics modeling of a manufacturing supply chain system", Simulation Modeling Practice and Theory, vol. 15, pp. 1338-1355, 2007.

[50] C. Eden, "Cognitive mapping and problem structuring for system dynamics model building", System Dynamics Review, vol. 10(2-3), pp. 257-276, 1994.

[51] S. Wolfram, Theory and application of cellular automata. World Scientific, 1986.

[52] A. Wuensche and M. J. Lesser, The global dynamics of cellular automata. Santa Fe Institute Studies in the Science of Complexity, Addison Wesley, 1992.

[53] P. Pal Chaudhuri, D. R. Chowdhury, S. Nandi, and S. Chatterjee, Additive cellular automata - Theory and applications, vol. 1. IEEE Computer Society Press, 1997.

[54] S. Wolfram, A new kind of science. Wolfram Media, 2002.

[55] L. Gray, "A mathematician looks at Wolfram's New Kind of Science, Notices of the American Mathematical Society, vol. 50(2), pp. 200-211.

[56] J. von Neumann, The theory of self-reproducing automata. Urbana: University of Illinois Press, 1966.

[57] A. Adamatzky, "Hierarchy of fuzzy cellular automata", Fuzzy Sets and Systems, vol. 62, pp. 167-174, 1994.

[58] B. Chopard and M. Droz, Cellular automata modeling of physical systems. Cambridge University Press, 1998.

[59] R. Hegselmann and A. Flache, "Understanding complex social dynamics: A plea for cellular automata based modeling", Journal of Artificial Societies and Social Simulation, vol. 1(3), June 1998.

[60] W. Li, N. H. Packard, and C. G. Langton, "Transition phenomena in cellular automata rule space", Physica D, vol. 45, pp.7-94, 1990.

[61] T. Toffoli and N. Margolus, Cellular automata machines. The MIT Press, 1987.

[62] S. Wolfram, "Computation theory of cellular automata", Commun. Math. Phys., vol. 96, pp. 15-57, 1984.

[63] S. Wolfram, "Universality and complexity in cellular automata", Physica D, vol. 10, pp. 1-35, 1984.

[64] S. A. Billings and Y. Yang, "Identification of probabilistic cellular automata", IEEE Trans. System, Man and Cybernetics, Part B, vol. 33(2), pp.1-12, 2002.

[65] O. Martin, A. M. Odlyzko, and S. Wolfram, "Algebraic properties of cellular automata", Commun. Math. Phys., vol. 93, pp. 219-258, 1984.

[66] G. Grinstein, C. Jayaprakash, and Y. He, "Statistical mechanics of probabilistic cellular automata", Phys. Rev. Lett., vol. 55, pp. 2527-2530, 1985.

[67] H. Gutowitz, "A hierarchical classification of CA", Physica D, vol. 45, pp. 136, 1990.

[68] G. Cattaneo, P. Flocchini, G. Mauri, C. Q. Vogliotti, and N. Santoro, "Cellular automata in fuzzy background", Physica D, vol. 105(1-3), pp. 105-120, June 1997.

[69] P. Flocchini, F. Geurts, A. Mingarelli, and N. Santoro, "Convergence and aperiodicity in fuzzy cellular automata: Revisiting rule 90", Physica D, vol. 142(1-2), pp. 20-28, August 2000.

[70] B. Kosko, Neural networks and fuzzy systems - A dynamical systems approach to machine intelligence. New Jersey: Prentice Hall, 1992.

[71] S.J. Russell and P. Norvig, Artificial intelligence: A modern approach. New Jersey: Prentice Hall, 1995.

[72] J.J. Hopfield, "Neural networks and physical system with emergent collective computational abilities", Proceedings of National Academy of Sciences, vol. 79, pp. 2554-2558, 1982.

[73] H. Ying, Fuzzy control and modeling: Analytical foundations and applications. Piscataway: IEEE Press, 2000.

[74] O. Castillo and P. Melin, "A new fuzzy inference system for reasoning with multiple differential equations for modeling complex dynamical systems", in Proceedings of CIMCA'99, IOS Press, Vienna, Austria, pp. 224-229, 1999.

[75] R.R. Yager, "On the construction of hierarchical fuzzy systems models", IEEE Trans. on Systems, Man and Cybernetics, vol. 28(1), pp. 55-66, 1998.

[76]   R. Axelrod, Structure of decision: the cognitive maps of political elites. New Jersey: Princeton University Press, 1976.
[77]   B. Kosko, "Fuzzy cognitive map", Int. J. Man-Machine Studies, vol. 24, pp. 65-75, 1986.
[78]   R. Taber, "Knowledge processing with fuzzy cognitive maps", Expert Systems with Applications, vol. 2(1), pp. 83-87, 1991.
[79]   W.R. Zhang and S. S. Chen, "A logical architecture for cognitive maps", in Proceedings of the IEEE International Conference on Neural Networks, 1988.
[80]   W.R. Zhang, S. S. Chen, and J.C. Besdek, "Pool2: a generic system for cognitive map development and decision analysis", IEEE Transactions on Systems, Man, and Cybernetics, vol. 19(1), pp. 31-39, 1989.
[81]   W.R. Zhang, S.S. Chen, W. Wang, and R.S. King, "A cognitive map based approach to the coordination of distributed cooperative agents", IEEE Trans. System, Man, and Cybernetics, vol. 22(1), pp.103-114, 1992.
[82]   M.A. Styblinski and B.D. Meyer, "Signal flow graphs vs fuzzy cognitive maps in application to qualitative circuit analysis", International Journal of Man-Machine Studies, vol. 35, pp. 175-186, 1991.
[83]   J.A. Dickerson and B. Kosko, "Fuzzy virtual worlds", AI Expert, pp. 25-31, July 1994.
[84]   K. Gotoh, J. Murakami, T. Yamaguchi and Y. Yamanaka, "Application of fuzzy cognitive maps to supporting for plant control", in Proceedings of SICE Joint Symposium of 15th Systems Symposium and 10th Knowledge Engineering Symposium, pp.99-104, 1989.
[85]   C. Stylios and P. Groumpos, "Application of fuzzy cognitive maps in large manufacturing systems", in Proceedings of the IFAC LSS'98, Rio, Patras, Greece, vol. 1, pp. 531-536, 1998.
[86]   C.E. Pelaez and J. B. Bowles, "Using fuzzy cognitive maps as a system model for failure models and effects analysis", Information Sciences, vol. 88, pp. 177-199, 1996.
[87]   C.E Pelaez and J. B. Bowles, "Applying fuzzy cognitive maps knowledge-representation to failure modes effects analysis", in Proceedings of Annual Reliability and Maintainability Symposium, pp. 450-455, 1995.
[88]   C.D. Stylios, V. C. Georgopoulos and P. P. Groumpos, "Applying fuzzy cognitive maps in supervisory control systems", in Proceedings of the European Symposium on Intelligent Techniques, Bari, Italy, March 1997, pp. 131-135.
[89]   L. Chun-Mei, "Using fuzzy cognitive map for system control, WSEAS Transactions on Systems, vol. 7(12), pp. 1504-1515, 2008.
[90]   E. Koulouriotis, I. Diakoulakis, D. Emiris, *et al.*, "Efficiently modeling and controlling complex dynamic systems using evolutionary fuzzy cognitive maps, Int. J. Computational Cognition, vol. 1(2), pp. 41-65, 2003.
[91]   J.R. Anderson, C.F. Boyle, A.T. Corbett, and M.W. Lewis, "Cognitive modeling and intelligent tutoring", Artificial Intelligence, vol. 42, pp. 7-49.
[92]   T. Kohonen, Self-organizing maps. Springer-Verlag, 2001.
[93]   M. Wooldridge, An introduction to multi-agent systems. Chichester: John Wiley and Sons, 2002.
[94]   G. Weiss, Multiagent systems: A modern approach to distributed artificial intelligence. Cambridge: The MIT Press, 2000.
[95]   M. Wooldridge and N. R. Jennings, "Intelligent agents: theory and practice, The Knowledge Engineering Review, vol. 10(2), pp. 115-152, 1995.
[96]   J.M. Epstein and R.L. Axtell, Growing artificial societies: Social science from the bottom up. Cambridge: MIT Press, 1996.
[97]   R. Axelrod, The complexity of cooperation: Agent-based models of competition and collaboration. Princeton: Princeton University Press, 1997.
[98]   C. Iglesias, M. Garijo Ayestaran, and J. Gonzalez, "A survey of agent-oriented methodologies", in Intelligent Agents V. Springer-Verlag, 1999.
[99]   L. Padgham and M. Winikoff, Developing intelligent agent systems: A practical guide. Chichester: John Wiley and Sons, 2004.
[100]  L. SeongKee, C. SungChan, L. HyengHo, and Y. ChanGon, "An agent-based system design for complex systems", WSEAS Transactions on Systems, vol. 5(9), pp. 2140-2146, 2006.
[101]  S. Lee, S. Cho, *et al.*, "A Layered agent oriented engineering approach for building complex systems, in Proceedings of the WSEAS International Conference on Systems, Athens, Greece, 2006, pp. 681-686.
[102]  J. Cuena and S. Ossowski, "Distributed models for decision support", in Weiss, Ed., Multi-Agent Systems - A Modern Approach to DAI. MIT Press, 1999.
[103]  M.V. Sokolova and A. Fernrandez-Caballero, "Agent-based decision making through intelligent knowledge discovery", in I. Lovrek, R.J. Howlett, L.C. Jain, Eds., KES 2008, Part III., LNCS (LNAI), vol. 5179, Springer, Heidelberg, 2008, pp. 709-715.

[104] B. Frankovič and V. Oravec, "Design of the agent-based intelligent control system", Acta Polytechnica Hungarica, vol. 2(2), pp 39-52, 2005.

[105] F. Zambonelli, N. Jennings, and M. Wooldridge, "Organizational abstractions for the analysis and design of multi-agent systems", in Ciancarini and Wooldridge, Eds., Agent-Oriented Software Engineering, Springer-Verlag, 2000, pp. 235-252.

[106] D. Xu, R. Volz, T. Ioerger, and J. Yen, "Modeling and verifying multi-agent behaviors using predicate/transition nets", in Proceedings of the International Conference on Software Engineering and Knowledge Engineering, 2002, pp. 193-200, ACM Press.

[107] H. L. Zhang, C. H. C. Leung, and G. K. Raikundalia, "Classification of intelligent agent network topologies and a new topological description language for agent networks," in Proceedings of the 4th IFIP International Federation for Information Processing, vol. 228, Springer, 2006, pp. 21-31.

[108] C. Castelfranchi, "Founding agent's autonomy on dependence theory", in Proceedings of the European Conference on Artificial Intelligence, Berlin, Germany, 2000, pp. 353-357.

[109] S.C. Bankes, "Tools and techniques for developing policies for complex and uncertain systems", Proceedings of National Academy of Sciences, vol. 99, suppl. 3, pp. 7263-7266, 2002.

[110] M. Schneider, E. Schneider, A. Kandel, G. Chew, "Automatic construction of FCMs", Fuzzy Sets and Systems, vol.93, pp.161-172, 1998.

[111] D. Harel, "Statecharts: A visual formalism for complex systems", Science of Computer Programming, vol. 45, pp. 231-274, 1987.

[112] J. Ryan and C. Heavey, "Development of a process modeling tool for simulation", Journal of Simulation, vol. 1, pp. 203-213, 2007.

[113] N. Boccara, Modeling complex systems. Graduate Texts in Contemporary Physics, New York: Springer, 2004.

[114] T. Dean and M. Wellman, Planning and control. Morgan Kaufmann, 1991.

[115] M. Klein and L. Methlie, Knowledge-based decision support systems. John Wiley & Sons, 1995.

[116] O. Balci and W. F. Ormsby, "Conceptual modeling for designing large-scale simulations", Journal of Simulation, vol. 1, pp. 175-186, 2007.

[117] L.A.N. Amaral and B. Uzzi, "Complex systems - A new paradigm for the integrative study of management, physical, and technological systems, Management Science, vol. 53(7), pp. 1033-1035, 2007.

[118] K. van der Heijden, Scenarios: The art of strategic conversation. New York: John Wiley and Sons, 1996.

[119] A.G. Bagdasaryan, "System theoretic viewpoint on modeling of complex systems: Design, synthesis, simulation, and control", in A. Zaharim, N. Mastorakis, and I. Gonos, Eds., Recent Advances in Computational Intelligence, Man-Machine Systems and Cybernetics, WSEAS Press, Stevens Point, 2008, pp. 248-253.

[120] Armen Bagdasaryan, "Discrete dynamic simulation models and technique for complex control systems", Simulation Modeling Practice and Theory, vol. 19(4), pp. 1061-1087, 2011.

[121] A.G. Bagdasaryan, Discrete modeling and analysis of complex dynamic systems in control mode. Moscow: Institute for Control Sciences RAS, 2005.

[122] A.G. Bagdasaryan, "Mathematical and computer tool of discrete dynamic modeling and analysis of complex systems in control loop", International Journal of Mathematical Models and Methods in Applied Sciences, vol. 2(1), pp. 82-95, 2008.

[123] A.G. Bagdasaryan, "Mathematical tool of discrete dynamic modeling of complex systems in control loop", in C. Long, S.H. Sohrab, G. Bognar, and L. Perlovsky, Eds., Recent Advances in Applied Mathematics, Mathematics and Computers in Science and Engineering: A Series of Reference Books and Textbooks, WSEAS Press, Stevens Point, March 2008, pp. 136-142.

[124] R.C. Thomas, "Qualitative models of complex systems", in Wharton Workshop on Complexity and Management, April 1999.

[125] A.G. Bagdasaryan, "System approach to synthesis, modeling and control of complex dynamical systems", WSEAS Trans. Systems and Control, vol. 4(2), pp. 77-87, 2009.

[126] A.G. Bagdasaryan and T.-h. Kim, "Dynamic simulation and synthesis technique for complex control systems", in D. Slesak, T.-h. Kim, A. Stoica, and Byeong-Ho Kim, Eds., Control and Automation, LNCS-CCIS, vol. 65, Springer-Verlag, Berlin-Heidelberg, 2009, pp. 15-27.

[127] H. Chen, Y. Zhu, K. Hu, X. He, "Hierarchical swarm model: A new approach to optimization", Discrete Dynamics in Nature and Society, vol. 2010, Article ID 379649, 30pp.

[128] A.G. Bagdasaryan, "A model of automated information system for solving control problems of large-scale systems", Control Sciences, no. 6, pp. 65-68, 2005.

[129] A.G. Bagdasaryan, "A general structure of information expert system for simulation and analysis of complex hierarchical systems in control loop", Control of Large-Scale Systems, vol. 21, pp. 58-70, 2008.

[130] G.S. Cumming and J. Collier, "Change and identity in complex systems", Ecology and Society, vol. 10(1), pp. 29-41, 2005.

[131] S. Wilson and C. Boyd, "Structured assessment of complex systems". ACPL-REPORT-12-2006-J101, April 2008, Aerospace Concepts Pty Ltd.

[132] J. Keppens and Q. Shen, "On compositional modeling", Knowledge Engineering Review, vol. 16, pp. 157-200, 2001.

[133] C.A. Knoblock, "Building a planner for information gathering: A report from the trenches", in Proceedings of the Third International Conference on Artificial Intelligence Planning Systems'AIPS96, Edinburgh, Scotland, 1996.

[134] H. Pattee, "Dynamic and linguistic models of complex systems", International Journal of General Systems, vol. 3, pp. 259-266, 1997.

[135] F. Bergenti and A. Ricci, "Three approaches to the coordination of multiagent systems", in Proceedings of the 2002 ACM Symposium on Applied Computing, 2002, pp. 367-372.

[136] J.J. Bryson, "Intelligence by design: Principles of modularity and coordination for engineering complex adaptive agents", Ph.D. Thesis, MIT, USA, 2001.

[137] R. Axelrod, The evolution of cooperation. Basic Books, 1984.

[138] N. Busi, P. Ciancarini, R. Gorrieri, and G. Zavattaro, "Coordination models: A guided tour", in A. Omicini, F. Zambonelli, M. Klusch, and R. Tolksdorf, Eds., Coordination o f Internet Agents: Models, Technologies, and Applications, 2001, pages 6-24.

[139] F. Schweitzer, J. Zimmermann, and H. Muhlenbein, "Coordination of decisions in a spatial agent model", Physica A, vol. 303 (1-2), pp. 189-216, 2002.

[140] E. Elliott and L.D. Kiel, "Exploring cooperation and competition using agent-based modeling", Proceedings of National Academy of Sciences, vol. 99, pp. 7193-7194, 2002.

[141] F. Atay and J. Jost, "On the emergence of complex systems on the basis of the coordination of complex behavior of their elements", Tech. Rep. 04-02-005, Santa Fe Institute for Studies of Complexity, Santa Fe, NM, USA, 2003, http://www.santafe.edu/media/workingpapers/04-02-005.pdf.

[142] F. Zambonelli, N.R. Jennings, and M.J. Wooldridge, "Organizational rules as an abstraction for the analysis and design of multiagent systems", International Journal of Software Engineering and Knowledge Engineering, vol. 11(4), pp. 303-328, 2001.

[143] K.M. Carley, "Computational organizational science and organizational engineering", Simulation Modeling Practice and Theory, vol. 10, pp. 253- 269, 2002.

[144] T. Malone and K. Crowston, "The Interdisciplinary study of coordination", ACM Computing Surveys, vol. 26(1), pp. 87-119, 1994.

[145] F. von Martial, Co-coordinating plans of autonomous agents. Springer-Verlag, 1992.

[146] S. Ossowski, Coordination in artificial agent societies. Social structure and its implications for autonomous problem-solving agents. Springer-Verlag, 1999.

[147] F. Kuhl, R. Wetherly, and J. Dahmann, Creating computer simulation systems: An introduction to the high level architecture. Prentice Hall, 1999.

[148] S. Haag, M. Cummings, D. McCubbrey, Management information systems for the information age. New York: McGraw-Hill, 2003.

[149] M.A. Azadeh, "Design of an intelligent simulation environment for manufacturing systems", in Proceedings of the Third International ICSC Congress on World Manufacturing , Symposium on Manufacturing Systems (ISMS'2001), Center for Integrated Manufacturing Studies, Rochester, New York, September 2001.

[150] M. Rao, T.S. Jiang and J.P. Tsai, "Integrated intelligent simulation environment", Simulation, vol. 54, pp. 291-295, 1990.

[151] S. Benjaafar, "Intelligent simulation for flexible manufacturing systems: An integrated approach", Computers and Industrial Engineering, vol. 22, pp. 297-311, 1992.

[152] Sakthivel and Agarwal, "Knowledge-based model construction for simulating information systems", Simulation, vol. 59, pp. 223-236, 1992.

[153] W.M. Sztrimbely and P.J. Weymouth, "Dynamic process plant simulation and scheduling: An expert system approach", Simulation, vol. 56, pp. 175-178, 1991.

[154] S.Y. Xu, Z. P. Jiang, *et al.*, "Control-theoretic results on dynamic decision making", WSEAS Transactions on Systems and Control, vol. 3(6), pp. 576-584, 2008.

[155]  D.J. Power, Decision support systems: Concepts and resources for managers. Westport: Quorum Books, 2002.
[156]  A.A.B. Pritsker, Decision support systems for engineers and scientists. London: Intl. Computer Graphics User Show, 1985.
[157]  R.H. Sprague and E.D. Carlson, Building effective decision support systems. Prentice-Hall, Englewood Cliffs, 1982.
[158]  M.S. Levin, Composite systems decisions. Springer, 2006.
[159]  R.L. Ackoff, "Management mis-information systems", Management Science, vol. 14(4), pp. 147-156, 1967.
[160]  D.M. Strong, "Decision support for exception handling and quality control in office operations", Decision Support Systems, vol. 8(3), pp. 217-227, 1992.
[161]  P.Y.K. Chau, "Decision support using traditional simulation and visual interactive simulation", Information and Decision Technologies, vol. 19, pp. 63-76, 1993.
[162]  R.J. Lempert, "A new decision sciences for complex systems", Proceedings of National Academy of Sciences, vol. 99, suppl. 3, pp. 7309-7313, 2002.
[163]  V.V.S. Sarma, "Decision making in complex systems", Systems Practice, vol. 7(4), pp. 399-407, 1994.
[164]  S.R. Watson and D.M. Buede Decision Synthesis: The Principles and Practices of Decision Analysis. Cambridge Univ. Press, New York, 1987.
[165]  S. French. Decision analysis and decision support systems. University of Manchester, 2000.
[166]  R.M. May, "Simple mathematical models with very complicated dynamics", Nature, vol. 261, pp. 459-467, 1976.